\pdfoutput=1 %
\PassOptionsToPackage{numbers,compress}{natbib}

\newif\ifarxiv
\arxivtrue

\ifarxiv
  \documentclass[twocolumn,10pt]{jmlr}
\else
  \documentclass[pmlr,twocolumn,10pt]{jmlr} %
\fi

\mlhtrack{proceedings}

\newif\iffinal
\finaltrue  %

\ifarxiv
    \jmlrproceedings{}{}
    \jmlrworkshop{Machine Learning for Health (ML4H) 2026}
\else\iffinal
    \ifmlhneedspmlr
      \jmlrvolume{346}
      \jmlryear{2026}
    \fi
    \ifmlhfindings \jmlrproceedings{}{ML4H 2026 - Findings Track}\fi
    \ifmlhdemo     \jmlrproceedings{}{ML4H 2026 - Demo Track}\fi
    \jmlrworkshop{Machine Learning for Health (ML4H) 2026}
\else
    \jmlrproceedings{}{Submitted to ML4H 2026: \mlhtrackname}
    \jmlrworkshop{Machine Learning for Health (ML4H) 2026}
\fi\fi

\usepackage[utf8]{inputenc} %
\usepackage[T1]{fontenc}    %
\usepackage{hyperref}       %
\usepackage{url}            %
\expandafter\def\expandafter\UrlBreaks\expandafter{\UrlBreaks\do\-} %
\usepackage{booktabs}       %
\usepackage{amsfonts}       %
\usepackage{nicefrac}       %
\usepackage{microtype}      %
\usepackage{xcolor}         %
\usepackage{graphicx}       %
\usepackage{bm}             %
\usepackage{amsmath}
\usepackage{amssymb}
\usepackage{multirow}
\usepackage{wrapfig}
\usepackage{xspace}
\usepackage{stfloats}    %

\usepackage{tikz}

\usepackage{tcolorbox}
\tcbuselibrary{skins, breakable, theorems}

\newtcolorbox{findingbox}{%
  enhanced, breakable,
  colback=white, colframe=black,
  boxrule=0.6pt, arc=4pt,
  left=6pt, right=6pt, top=5pt, bottom=5pt,
  before skip=8pt, after skip=8pt,
}

\newtcbtheorem[auto counter, number within=section]{prompt}%
  {Prompt}%
  {%
    enhanced, breakable,
    colback=gray!5, colframe=black!75,
    fonttitle=\bfseries,
    boxrule=0.5pt,
    left=6pt, right=6pt, top=4pt, bottom=4pt,
    before skip=8pt, after skip=8pt,
  }%
  {prompt}

\definecolor{purple}{RGB}{102,51,153}
\newcommand{\ph}[1]{\textcolor{purple}{\texttt{\textless{}#1\textgreater{}}}}

\newcommand{\oursfull}{PatientsWithPersonality\xspace}
\newcommand{\ours}{PWP\xspace}

\newcommand{\Fmeta}{\Phi_{\mathrm{meta}}}               %
\newcommand{\Fconv}{\Phi_{\mathrm{conv}}}               %

\newcommand{\mistral}{\texttt{Ministral-3-14B}\xspace}          %
\newcommand{\gemini}{\texttt{Gemini-3.1-Flash-Lite}\xspace}     %
\newcommand{\geminiflash}{\texttt{Gemini-3.5-Flash}\xspace}          %
\newcommand{\geminipro}{\texttt{Gemini-3.1-Pro}\xspace}              %
\newcommand{\claude}{\texttt{Claude-Sonnet-5.5}\xspace}           %

\newcommand{\qwen}{\texttt{Qwen3-Embedding-0.6B}\xspace}        %

\newcommand{\case}{\mathbf{x}}                           %
\newcommand{\caseentry}{x}                               %
\newcommand{\caseof}[1]{\mathbf{x}_{#1}}                 %
\newcommand{\casemask}{\tilde{\mathbf{x}}}               %

\newcommand{\Fall}{\mathcal{F}}                          %
\newcommand{\Fper}{\mathcal{F}_{\mathrm{per}}}           %
\newcommand{\Flife}{\mathcal{F}_{\mathrm{life}}}         %
\newcommand{\Fmed}{\mathcal{F}_{\mathrm{med}}}           %

\newcommand{\persona}{\bm{\theta}}                       %
\newcommand{\persH}{h}                                   %
\newcommand{\persE}{e}                                   %
\newcommand{\persX}{x}                                   %
\newcommand{\persA}{a}                                   %
\newcommand{\persC}{c}                                   %
\newcommand{\persO}{o}                                   %
\newcommand{\persL}{l}                                   %
\newcommand{\axisdesc}{\boldsymbol{d}_{\persona}}

\newcommand{\sbelief}{s_{\mathrm{belief}}}               %
\newcommand{\semotion}{s_{\mathrm{emotion}}}             %
\newcommand{\stangent}{s_{\mathrm{tangent}}}
\newcommand{\latentrole}{r}                              %

\newcommand{\question}{q}                                %
\newcommand{\answer}{y}                                  %
\newcommand{\relevant}{R}                                %
\newcommand{\history}{H}                                 %

\newcommand{\convtype}{\rho}                             %
\newcommand{\answerformrate}{\beta}                      %
\newcommand{\turnform}{\phi}                             %

\newcommand{\pHdown}{p_{\texttt{H\_DOWNPLAY}}}
\newcommand{\pCfuzzy}{p_{\texttt{C\_FUZZY}}}
\newcommand{\platent}{p_{\texttt{LATENT\_ROLE}}}

\newcommand{\pclassform}{p_{\texttt{CLASS\_FORM}}}
\newcommand{\psys}{p_{\texttt{SYS}}}

\usepackage{siunitx}

\usepackage[switch]{lineno}

\newcommand{\equal}[1]{{\hypersetup{linkcolor=black}\thanks{#1}}}

\title[Patients With Personality]{Patients With Personality: Realistic Patient Simulation through Controlled Diversity and Selective Disclosure}

\author[Schlager et al.]{%
\Name{Moritz Schlager\nametag{$^{1,2}$\thanks{Corresponding author}}} \Email{moritz.schlager@tum.de}\\
\Name{Friederike Jungmann\nametag{$^{1,3}$}}\\
\Name{Samuel Schmidgall\nametag{$^{4}$}}\\
\Name{Philipp Raffler\nametag{$^{5}$}}\\
\Name{Franziska Hartl\nametag{$^{5}$}}\\
\Name{Eva Wende\nametag{$^{3}$}}\\
\Name{Paula Roßmüller\nametag{$^{5}$}}\\
\Name{Conrad Ketzer\nametag{$^{6}$}}\\
\Name{Matthew Thompson\nametag{$^{7}$}}\\
\Name{Justin Anthony Chen\nametag{$^{7}$}}\\
\Name{Christina Chen\nametag{$^{7}$}}\\
\Name{Avinatan Hassidim\nametag{$^{7}$}}\\
\Name{Dale R.\ Webster\nametag{$^{7}$}}\\
\Name{Yossi Matias\nametag{$^{7}$}}\\
\Name{Yun Liu\nametag{$^{7}$}}\\
\Name{Daniel Rueckert\nametag{$^{1,2,8}$}}\\
\Name{Mike Schaekermann\nametag{$^{7}$\equal{Equal supervision}}} \Email{mikeshake@google.com}\\
\Name{Paul Hager\nametag{$^{1,9}$\footnotemark[2]}} \Email{paul.hager@tum.de}\\
\addr $^{1}$Chair for AI in Healthcare and Medicine, Technical University of Munich (TUM) and TUM University Hospital, Munich, Germany;
$^{2}$Munich Center for Machine Learning (MCML), Munich, Germany;
$^{3}$Department of Radiology, TUM University Hospital, Munich, Germany;
$^{4}$Google DeepMind, Mountain View, CA, USA;
$^{5}$Department of Neuroradiology, TUM University Hospital, Munich, Germany;
$^{6}$Department of Trauma Surgery, TUM University Hospital, Munich, Germany;
$^{7}$Google Research, Mountain View, CA, USA;
$^{8}$Imperial College London, London, UK;
$^{9}$Virtual Diagnostics Unit, QuantCo, Cambridge, MA, USA
}
\jmlrauthors{M.~Schlager et al}

\iffinal\else\linenumbers\fi %

\begin{document}

\maketitle

\begin{figure}[p]
\centering
\includegraphics[width=\linewidth]{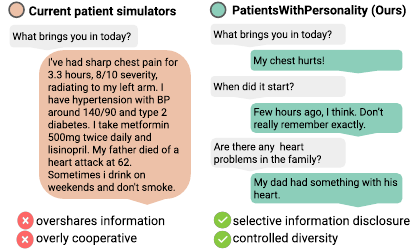}
\caption{Our framework resolves failure modes of current patient simulators by selectively controlling information disclosure rate while allowing for diverse patient personalities.}
\label{fig:motivation}
\end{figure}

\ifmlhdemo\else
\begin{abstract}
Realistic patient simulation is a key requirement for developing and testing clinical LLMs at scale without costly user studies. Existing simulators lack realism and controllability: they overshare information and miss the variability of real patient behavior. Here, we introduce \textit{\oursfull (\ours)}, a patient simulation framework that generates realistic yet diverse patient responses through explicit personality parametrization over a latent patient state. Grounded in the six-dimensional HEXACO personality model, it gives fine-grained control over conversational style, cooperativeness, and information disclosure. We evaluate \ours in the first direct comparison of different patient simulators under a shared protocol, anchored in transcripts of real physician--patient encounters. Clinicians judge \ours conversations as realistic at least as often as the recorded encounters themselves and far more often than prior simulators, and almost never flag them as ``too informative''. Our simulator also reproduces the disclosure pacing of real patients, and its HEXACO-conditioned personality traits can be recovered by clinicians and an autorater. \ours further shows a substantially wider behavioral range than the closest baseline. In a downstream benchmark, LLMs interviewing patients simulated by \ours perform closest to how they perform on the real encounters, whereas prior simulators make the task easier. \ours thus provides a realistic, steerable, and diverse patient population for improving and benchmarking clinical LLMs.
\end{abstract}
\fi

\ifmlhneedsstatements
\paragraph*{Data and Code Availability}
All datasets used in this work are publicly available, and source details
are given in Appendix~\ref{app:datasets}. The complete implementation of our framework and all experiments is available at \url{https://github.com/mo374z/PatientsWithPersonality}.

\paragraph*{Institutional Review Board (IRB)}
This research does not require IRB approval, as it uses only publicly available, de-identified datasets.
\fi

\section{Introduction}

Benchmarking large language models (LLMs) for use in the high-stakes medical domain guides model selection and informs deployment decisions. Benchmarking is most useful when models are assessed under conditions that match clinical reality~\citep{singhal_large_2023, tang_evaluating_2023}. Exam-style question-answering benchmarks are increasingly saturated and only moderately predictive of clinical capability~\citep{hager_evaluation_2024, kim_questioning_2025, wu_medarena_2026}, so they are being supplemented by interactive and agentic evaluations that match actual clinical scenarios more closely, thus narrowing the knowledge-practice gap~\citep{gong_knowledge-practice_2025,schmidgall_agentclinic_2026,johri_evaluation_2025}. Such simulations are also cheaper than repeated human studies, enabling much tighter development cycles and accelerating the rate of progress.

A foundational building block of these interactive benchmarks is a \emph{simulated patient}: an LLM-driven agent that the clinician AI under evaluation interacts with. Although the patient simulator itself is rarely the object of study, its realism determines whether automated, scalable evaluation can stand in for costly human studies. However, unrealistic simulated patients can distort those results. \emph{ResidencyRL} trains clinician agents against adversarial LLM patients~\citep{lievin_residencyrl_2026}, where the reward is only as informative as the patient it is computed against. In the clinical benchmarking domain, \citet{bean_reliability_2026} found in a randomized study of more than 1,000 participants that simulated patients substantially inflated model performance and correlated only weakly with results from real patients. Benchmarks against idealized patients thus overestimate clinical capability by a margin that depends on the simulator's fidelity. 

Two recurring failure modes explain the discrepancies mentioned above: first, simulators tend to overshare, volunteering information that was not explicitly asked for, and flesh out underspecified facts with fictitious content~\citep{holderried_generative_2024}. Second, they produce uniform and cooperative patients, lacking the imperfect recall and personality-driven variation that characterize real patient encounters~\citep[][cf.\ Figure~\ref{fig:motivation}]{cook_virtual_2025,jiang_multi-stage_2026,hosseinpourkhoshkbari_spot_2025}. \citet{matos_complexities_2026} showed that by replaying five distinct patient personae through four LLMs, these failures measurably shift urgency assessments.

\paragraph{Our contributions.}
To address both failure modes, we introduce \emph{\oursfull (\ours)}, a patient simulation framework that decouples behavioral personality modeling from information disclosure. Specifically, we contribute:
\begin{itemize}
    \item The \textit{\ours} framework featuring a flexible, HEXACO-grounded personality configuration and dynamic information control through a query-conditioned disclosure mask;
    \item A physician-aligned simulation protocol that replays the physician turns of real encounters, enabling the first direct comparison of patient simulators on common ground and realism evaluation against real rather than actor-portrayed patients;
    \item A clinician study showing that \textit{\ours} is judged as realistic as the recorded encounters, matches human disclosure pacing, and produces personas that are both recoverable and more diverse than the closest parametrizable baseline;
    \item A downstream contextual-care benchmark built from expert annotations of the real encounters, on which \ours reproduces the task difficulty of real patients across three LLMs.
\end{itemize}

\section{Background \& Related Work}\label{sec:related}

\paragraph{LLM-based patient simulators.}
LLM-driven patient simulators have emerged as a scalable alternative to real humans across a range of clinical applications. They serve as the patient side of multi-agent benchmarks \citep{schmidgall_agentclinic_2026, johri_evaluation_2025, fan_ai_2025, fajardo_medpi_2025, srirag_triagesim_2026} or replace human patient actors in medical education \citep{holderried_generative_2024,cook_virtual_2025,du_llms_2025,wu_simulation_2026}. The use of these simulators allows for the probing of more realistic multi-turn communication settings that static tasks cannot capture \citep{liao_automatic_2024}.

Existing simulators can be separated into two broad families. The first are lightweight, solely prompt-based simulators that are typically components of a larger agentic benchmark rather than artifacts in their own right. \emph{CRAFT-MD} turns case vignettes into simulated doctor-patient dialogues to evaluate multi-turn reasoning and history taking~\citep{johri_evaluation_2025}. \emph{AgentClinic} embeds a prompt-driven patient agent inside a multi-agent, OSCE (Objective Structured Clinical Examination)-like diagnostic benchmark and adds cognitive and social bias perturbations to the patient agent~\citep{schmidgall_agentclinic_2026}. \emph{PatientAgentBench}~\citep{vatanparvar_patientagentbench_2026} and \emph{EasyMED}~\citep{zhang_human_2026} embed comparable patient agents in patient-facing and multi-agent consultation systems, respectively. In each case, the patient is a single behavioral prompt on top of a case description limiting steerability on an individual level.

The second family consists of standalone simulator frameworks that expose explicit control over factors such as disclosure or persona. \emph{StateAwarePatientSimulator (SAPS)} introduces inquiry-conditioned disclosure through a state-machine patient~\citep{liao_automatic_2024}, \emph{PatientSim} conditions the simulator on four independent axes: personality, CEFR language proficiency, medical history recall, and cognitive confusion~\citep{kyung_patientsim_2025}, and \emph{VeriSim} injects six clinically grounded noise dimensions (e.g., recall gaps, low health literacy, stigma-driven non-disclosure) at five severity levels while a UMLS-based verifier guards factual validity~\citep{mansouri_verisim_2026}. Another approach models clinical content, emotion, conversational strategy, and communication style as separate components~\citep{matos_complexities_2026}.

These frameworks establish that decoupled control and explicit disclosure mechanisms improve simulation quality, but their control axes remain largely prompted behavioral labels, and realism is judged by comparison to actor-portrayed encounters~\citep{hosseinpourkhoshkbari_spot_2025} or consumer chatbot logs~\citep{matos_complexities_2026} rather than to transcripts of real physician--patient visits. \citet{mansouri_verisim_2026} show that injecting communication noise lowers the diagnostic accuracy of tested LLMs and lengthens consultations. Crucially, they never performed a blind evaluation of the realism of their simulated patients, mainly checking for medical hallucinations. No prior work has compared representative simulators from both families in the clinical consultation setting under a shared evaluation spanning realism, disclosure, and personality fidelity.

\paragraph{Modeling personality.}
Patients differ not only in symptoms but in how they behave during an encounter. The HEXACO model defines personality structure through six dimensions derived from cross-linguistic lexical studies~\citep{ashton_empirical_2007}. Its axes capture truthfulness versus deceptiveness (\emph{Honesty-Humility}), emotional stability versus distress (\emph{Emotionality}), reserve versus talkativeness (\emph{eXtraversion}), cooperativeness versus skepticism (\emph{Agreeableness}), diligence versus forgetfulness (\emph{Conscientiousness}), and curiosity versus conventional thinking (\emph{Openness}). Relative to the Big Five~\citep{roccas_big_2002}, HEXACO adds a distinct \emph{Honesty-Humility} axis and separates \emph{Emotionality} from \emph{Agreeableness}. For a patient simulator, \emph{Honesty-Humility} anchors the disclosure of sensitive information and \emph{Emotionality} isolates distress expression. Prior work has shown that the six axes are empirically linked to health-relevant behavior~\citep{pletzer_who_2024}, and HEXACO-conditioned prompting systematically shifts LLM generation behavior~\citep{wang_exploring_2025}.

\section{Simulating Realistic Patients}\label{sec:framework}

\begin{figure*}[t]
\centering
\includegraphics[width=\linewidth]{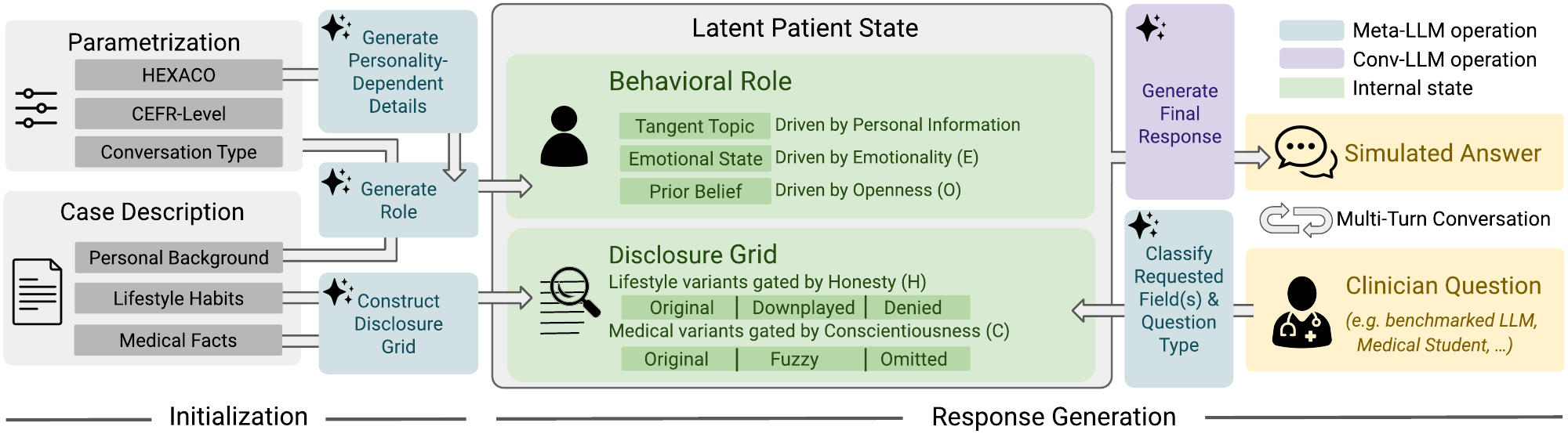}
\caption{Overview of the \emph{\ours} framework. During \textbf{initialization}, the personality parametrization and structured case description are transformed into a latent patient state by a meta-LLM. During \textbf{response generation}, the content and type of the incoming question then determine the information used to generate the response by the conversational LLM. The answer is returned to the external system, closing the multi-turn interaction loop.}
\label{fig:framework}
\end{figure*}

We introduce \textit{\ours}, a framework for simulating realistic and varied patient dialogue with a configurable behavioral personality and explicit control over the underlying information at each turn. The framework consists of two stages, shown in Figure~\ref{fig:framework}. During \emph{initialization}, a structured case description and a parametrization are transformed into a latent patient state comprising (i)~a \emph{behavioral role} that the simulator adheres to throughout the dialogue, and (ii)~a \emph{disclosure grid} that controls how facts are revealed across turns.
Clinical and lifestyle facts that were not requested are excluded from the generation prompt entirely; only the personal background of the role card is always visible. Questions can stem from any external system; in a standard benchmarking setup, this would be the clinician AI under evaluation.

\paragraph{Case description.}
A simulated patient is grounded in a structured case description $\case = (\caseentry_1,\dots,\caseentry_m)$, where each entry $\caseentry_i$ is a single fact identified by a named field. A field may carry multiple facts (e.g., \texttt{chiefcomplaint}: [high blood pressure; palpitations]). The set of fields $\Fall$ is partitioned into three disjoint semantic subgroups. We differentiate between personal background ($\Fper$), sensitive lifestyle habits ($\Flife$), and clinical facts ($\Fmed$). For a subset $S \subseteq \Fall$, $\caseof{S}$ denotes the corresponding sub-vector of facts. The full list of fields is given in Appendix~\ref{app:case_fields}.

\paragraph{Personality parametrization.}
Patient behavior is controlled by a parametrization vector
\[
\persona = (\persH,\persE,\persX,\persA,\persC,\persO,\persL) \in \{1,2,3\}^7,
\]
where the first six components are the axes of the HEXACO model of personality structure (Honesty-Humility, Emotionality, eXtraversion, Agreeableness, Conscientiousness, Openness), and $\persL$ is a language-proficiency level (CEFR). The HEXACO levels index how strongly the patient expresses the behaviorally challenging pole of each axis.

The HEXACO axes are mapped to simulator mechanisms by their behavioral connotation. Honesty-Humility and Conscientiousness are directly tied to disclosure: low honesty implies concealment of socially undesirable habits and low conscientiousness implies imprecise recall of clinical detail, so these two axes gate the disclosure of lifestyle and medical fields, respectively. Emotionality, eXtraversion, and Openness shape \emph{how} patients communicate rather than \emph{what} they disclose; we realize them by injecting case-specific content (distress, tangent topics, prior beliefs) into the behavioral role. Agreeableness only influences style.
We additionally define a conversation type $\convtype \in \{\text{written}, \text{spoken}\}$, since the structure of the simulated turns heavily depends on the setting. \emph{written} patients mimic asynchronous chat-based encounters, whereas \emph{spoken} patients answer as if speaking aloud, using contractions, occasional fillers, or self-corrections, which makes the simulation feel like the transcript of an in-person encounter.

\paragraph{LLM components.}
The framework uses a meta-LLM $\Fmeta$ for structured internal operations (classification, latent-variable and fact-variant generation) and a conversational LLM $\Fconv$ to generate the patient utterances. Each operation uses a dedicated prompt $p_\bullet$ (cf. Appendix~\ref{app:prompts}). The notation $\Fmeta(\cdot\,;\,p_\bullet)$ denotes invoking $\Fmeta$ with that prompt.

\paragraph{Behavioral role construction.}
The behavioral role is assembled from static and dynamic components. The personal background $\caseof{\Fper}$ and CEFR level $\persL$ are used verbatim. Each HEXACO axis contributes a single-sentence descriptor selected by its level in $\persona$:
\[
\axisdesc \triangleq \bigl(d_H(\persH),\,d_E(\persE),\,d_X(\persX),\,d_A(\persA),\,d_C(\persC),\,d_O(\persO)\bigr),
\]
where each $d_\bullet(\cdot)$ is a lookup into a fixed table of three short descriptions per axis (cf. Appendix Table~\ref{tab:prompts-hexaco}). For Emotionality, eXtraversion, and Openness set to the highest level, we additionally generate dynamic, case-specific content that a static descriptor cannot provide. For $\persE=3$, we generate a distress state $\semotion$ that affects communication style without altering clinical facts. For $\persX=3$, we draw a profile-consistent tangent topic $\stangent$ from a predefined, age-conditioned list. For $\persO=3$, we generate a plausible prior belief $\sbelief$ representing a self-diagnosis or lay explanation. We denote $\mathbf{s} = [\sbelief, \semotion, \stangent]$ for brevity.

All of these components are then synthesized into a coherent initial behavioral role description:

\[
\latentrole = \Fmeta\bigl(\caseof{\Fper},\,\persL,\,\axisdesc,\,\mathbf{s} \,;\, \platent\bigr).
\]

Because the meta-prompt synthesizing the latent role $\latentrole$ strongly affects downstream behavior~\citep{lu_fantastically_2022,zhou_large_2022,sadjoli_optimization_2025}, we optimize $\platent$ against a penalty measuring information and personality control on encounters disjoint from the main studies. The resulting prompt gives clear instructions around different dimensions of interaction and improves the tuning metric on held-out cases (cf. Appendix~\ref{app:metaprompt}).

\paragraph{Disclosure grid.}
Alongside the behavioral role, initialization sets up a disclosure grid that controls how case information is surfaced to $\Fconv$. Lifestyle and medical information are gated independently. For each atomic fact $\caseentry_i$ in either subgroup, we construct different variants, concretely:
\begin{itemize}
\item \textbf{Lifestyle facts ($\Flife$)}: (\emph{original}, \emph{downplayed}, \emph{denied}), where the downplayed variant is generated by $\Fmeta(\caseentry_i \,;\, \pHdown)$ and the denial is a fixed phrase per field (e.g., ``not drinking alcohol'').
\item \textbf{Medical facts ($\Fmed$)}: (\emph{original}, \emph{fuzzy}, \emph{omitted}), where the fuzzy variant is generated by $\Fmeta(\caseentry_i \,;\, \pCfuzzy)$ and the omitted variant masks the fact entirely from the prompt.
\end{itemize}

\begin{table}[t]
\centering
\small
\caption{Overview of information disclosure at different levels of personality traits $\persH$ and $\persC$.}
\begin{tabular}{lll}
\toprule
\textbf{Fields} & \textbf{Level} & \textbf{State and dynamics} \\
\midrule
\multirow{3}{*}{$\Flife$}
& $\persH=1$ & \emph{original}\\
& $\persH=2$ & $\sim \text{Unif} \{\emph{original}, \emph{downplayed}\}$\\
& $\persH=3$ & $\sim \text{Unif} \{\emph{downplayed}, \emph{denied}\}$\\
\midrule
\multirow{3}{*}{$\Fmed$}
& $\persC=1$ & \emph{original} on first targeted request \\
& $\persC=2$ & \emph{fuzzy} $\to$ \emph{original}\\
& $\persC=3$ &  \emph{omitted} $\to$ \emph{fuzzy} $\to$ \emph{original} \\
\bottomrule
\end{tabular}
\label{tab:disclosure}
\end{table}

Which variant is initially active is determined by the gating personality level, summarized in Table~\ref{tab:disclosure}. The state of the $\Flife$ fields is fixed after initialization (sampled only once per atomic fact), since Honesty-Humility is a stable trait, so a patient who downplays their alcohol intake on the first request keeps doing so. The fields gated by $\persC$ advance one variant on each targeted request, because recall depends on how specifically the patient is prompted, so targeted follow-ups progressively unmask medical information. Unlike the inquiry-conditioned implementation of SAPS~\citep{liao_automatic_2024}, this unmasking is progressive and gated by a personality trait, so repeated targeted questioning recovers a fact, not just a single disclosure decision.

\paragraph{Response generation.}
At turn $t$ of the simulated dialogue, \ours receives a question $\question_t$. First, a meta-operation classifies the question type $\turnform_t \in \{\text{polar}, \text{open}, \text{statement}\}$ and determines which fields $\relevant_t \subseteq \Flife \cup \Fmed$ are required to answer it:
\[
\bigl(\relevant_t,\,\turnform_t\bigr) = \Fmeta\bigl(\question_t,\,\Flife \cup \Fmed \,;\, \pclassform\bigr).
\]
Only fields in $\relevant_t$ become candidates for inclusion in the generation prompt; gated fields not in $\relevant_t$ are excluded entirely, while the personal background $\caseof{\Fper}$ stays part of the behavioral role throughout. For each candidate field, the disclosure grid determines which variant is shown. Fields in $\Flife \cap \relevant_t$ retain the variant fixed at initialization. Fields in $\Fmed \cap \relevant_t$ are progressively revealed on each request (cf. Table~\ref{tab:disclosure}). We denote the final view of the case description at turn $t$ as $\casemask_t$.
The question type $\turnform_t$ sets whether $\question_t$ is answerable with yes or no, asks for an elaborated answer, or is an acknowledgment or explanation by the physician requiring no direct response. With probability $\answerformrate$, the form of $\answer_t$ is aligned to $\turnform_t$ by adding a corresponding instruction to $\psys$, so that a polar question is often answered concisely, but by chance leads to a longer elaboration.

Finally, the conversational model generates the patient utterance, conditioned on the latent role, the currently accessible information, the conversation history, and the incoming question:
\[
\answer_t = \Fconv\bigl(\latentrole,\,\casemask_t,\,\history_{t-1},\,\question_t \,;\, \psys\bigr).
\]

\section{Experimental Setup}\label{sec:setup}

We evaluate \emph{\ours} in three studies: the \emph{realism study} (Section~\ref{sec:results_default}) compares the simulators under their default configurations, isolating realism from personality-specific effects. The \emph{extreme-personality study} (Section~\ref{sec:results_personality}) instantiates the framework at the extreme of each HEXACO axis to test whether configured traits are recoverable from simulated transcripts and whether behavior actually varies across configurations. The \emph{downstream benchmark} (Section~\ref{sec:results_downstream}) tests whether frontier LLMs interviewing the simulated patients perform as they do on the real encounters. In our setup, we use two distinct LLMs to operate our framework: as meta-LLM $\Fmeta$, for internal operations such as downplaying atomic facts or generating $\latentrole$, we use the lightweight \mistral{}, and as conversational LLM $\Fconv$, for utterance generation, we use the more capable \gemini{}. We also use the latter as the autorater to scale the study-based experiments reported below. %

\paragraph{Dataset.}
All experiments are based on encounters from a dataset of the Veterans Health Administration (VHA)~\citep{vha_2022}, a public release of de-identified primary-care transcripts between physicians and real patients. Unlike the OSCE encounters role-played by actors used in prior realism evaluations~\citep{hosseinpourkhoshkbari_spot_2025}, these transcripts record routine visits of real patients. Because the dataset only contains raw transcripts without accompanying structured patient profiles, we derive the case description $\case$ for each retained encounter directly from its transcript. Dataset construction, filtering criteria, and exact identifiers of the cases used are given in Appendix~\ref{app:datasets}.

\paragraph{Physician-aligned protocol.}
To enable a side-by-side comparison across simulators, we run all conversations under a shared \emph{physician-aligned} protocol: from the utterances in our dataset, the physician side is treated as a fixed track, and at every turn the simulator under evaluation produces the patient response in place of the real patient~\citep{hosseinpourkhoshkbari_spot_2025}. This removes variance from physician-side behavior while preserving the natural turn structure of an authentic encounter. Because a simulated patient response can deviate from the recorded one, making the fixed physician answer appear unsuitable, each subsequent physician turn is realigned to the preceding simulated utterance through a minimal edit using \gemini{} (e.g., answering a question before proceeding as in the transcript).

\paragraph{Clinician study.}
To judge the realism of the simulators, we recruited nine resident clinicians with 1--5 years of working experience to evaluate transcripts. Annotation was carried out through a custom labeling interface in which each clinician judged the same pool of 20 clinical cases. Every case is shown as three side-by-side conversations that are rated independently, yielding $540$ labeled encounters in total (cf. Appendix~\ref{app:doctor_study_details}). We compare \ours against the two leading representatives of each simulator family, namely AgentClinic (prompt-based) and PatientSim (explicitly parametrizable), which keeps the per-simulator sample sizes in the clinician study sufficiently high. We contrast the simulators against two human-based controls. \emph{Human Reference} denotes the original answers from the recorded conversation between physicians and real patients, while \emph{Human Rephrase} passes those same recorded answers through an LLM paraphrase. The latter isolates how the surface form of LLM-generated text is perceived when its content is strictly human.

\section{Results}\label{sec:results}

\subsection{Realism Study}\label{sec:results_default}

\begin{figure}[t]
\centering
\includegraphics[width=\linewidth]{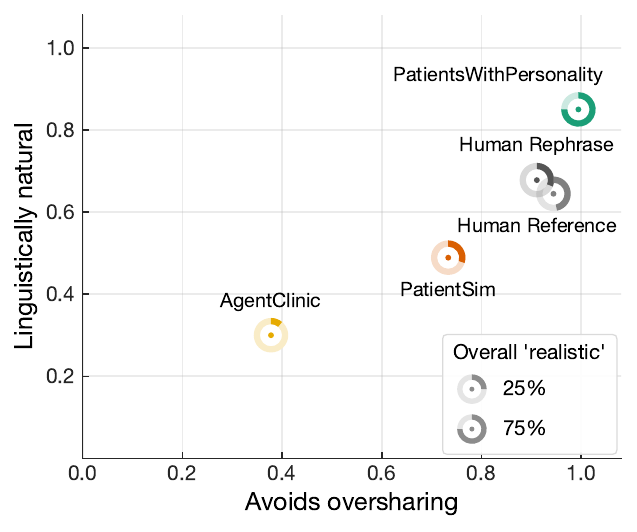}
\caption{Realism of different simulators judged by clinicians, displayed with a gauge around each marker showing the fraction of conversations judged as `real'. The axes show the complement of the fraction of conversations labeled `not real' for being \textit{too informative} (x) or \textit{linguistically unnatural} (y).}
\label{fig:reason_scatter}
\end{figure}

\paragraph{Clinician judgments of realism.}
For every conversation presented, clinicians determine whether it is real and, when it is not, give at least one reason. On overall realism, \ours is judged realistic $75.0\%$ of the time, ahead of the real human encounters ($46.7\%$), the rephrased transcripts ($31.1\%$), PatientSim ($30.0\%$), and AgentClinic ($11.1\%$). A mixed-effects logistic regression on the binary target, with crossed random intercepts for rater and case, sets \ours apart from every other patient type (all $p<10^{-4}$). In a head-to-head within cases, clinicians preferred \ours to the two other patient simulators in $114$ of the $180$ pairings, with the others preferred only $16$ times. To locate the failure modes behind these labels, Figure~\ref{fig:reason_scatter} plots the overall perceived realism against the two dominant rejection reasons, \emph{too informative} and \emph{linguistically unnatural}. \ours sits with the two human references in the top-right corner: it is judged realistic most often and flagged least for being either \emph{too informative} ($1/180$ cases against $24/90$ for PatientSim and $56/90$ for AgentClinic) or \emph{linguistically unnatural}. More details, including the rates for all rejection reasons and inter-rater reliability, are reported in Appendix~\ref{app:clinician_study_extended_results}.

\begin{figure*}[t]
\centering
\includegraphics[width=\linewidth]{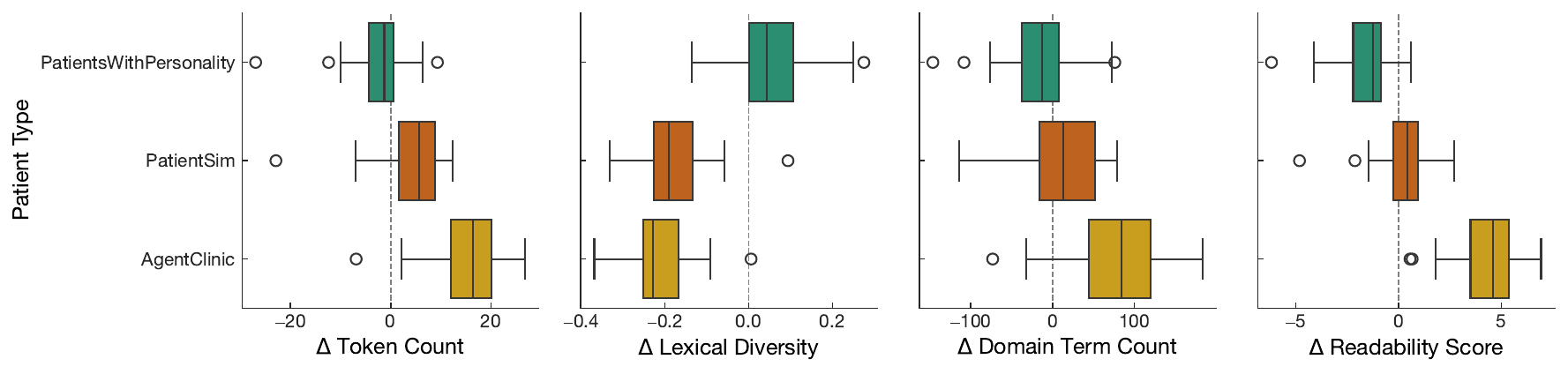}
\caption{Per-conversation differences ($\text{sim}-\text{real}$) between each simulator and the real transcript on four dialogue metrics, over the physician-aligned conversations. Zero marks agreement with the transcript.}
\label{fig:sim_comp_metrics}
\end{figure*}

\paragraph{Quantitative dialogue metrics.}
The physician-aligned protocol allows us to pair every simulated patient turn with its recorded human counterpart. As objective comparison extending on the clinician study results, we report the per-conversation signed difference to the Human Reference on four dialogue metrics in Figure~\ref{fig:sim_comp_metrics}. Values near zero indicate agreement with the recorded encounter. The metrics are the \emph{token count}, capturing the verbosity of the simulator; \emph{lexical diversity}, where lower values indicate more repetitive vocabulary; \emph{domain term count}, capturing how much clinical terminology the patient uses; and \emph{readability}~\citep{kincaid_derivation_1975}, where higher values indicate more complex speech. \ours and PatientSim both stay close to the reference across all four metrics, with \ours clearly closest to the Human Reference on response length and lexical diversity. AgentClinic deviates furthest, generating more lexically uniform, jargon-heavy, and harder-to-read utterances. The same pattern holds on $22$ multi-specialty encounters of ACI-Bench~\citep{yim_aci-bench_2023}, a population that is $77\%$ female and aged 22--100 (Appendix~\ref{app:aci}).

\paragraph{Disclosure behavior.}

\begin{figure}[t]
\centering
\includegraphics[width=\linewidth]{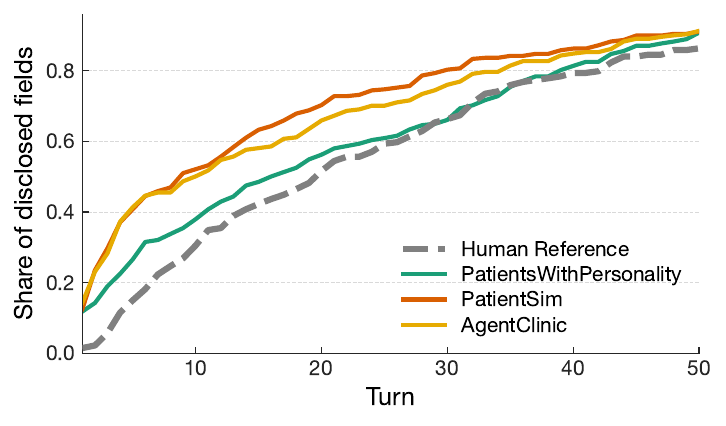}
\caption{Cumulative share of disclosed information during a conversation. Each conversation is normalized by the fields it eventually discloses before averaging.}
\label{fig:disclosure_pacing}
\end{figure}

To measure information-sharing behavior, we mark a case field as \emph{disclosed} at the first turn at which the simulated utterance contains content matching that field. Figure~\ref{fig:disclosure_pacing} shows that under default configurations, \ours tracks the pacing of the Human Reference closely, while PatientSim and AgentClinic front-load disclosure and diverge from the human curve. In absolute terms, both the Human Reference and \ours disclose a lower overall share of fields by the end of the conversation ($47.9\%$ and $42.1\%$, respectively) than PatientSim ($51.3\%$) and AgentClinic ($52.3\%$). Importantly, disclosure in \ours is configured through $\persona$, so a benchmark that wants a more forthcoming patient can elicit this, whereas AgentClinic and PatientSim expose no equivalent control. More metrics on disclosure are given in Appendix~\ref{app:info_disclose_extended}.

\subsection{Extreme-Personality Study}\label{sec:results_personality}

To isolate extreme personality behavior on each axis, we instantiate six \ours configurations, each with exactly one HEXACO axis at its extreme ($3$), while the remaining five are held at level $1$, and pair each with the PatientSim parametrization of the closest behavioral analogue (cf. Appendix~\ref{app:extreme_study}).

\begin{table*}[t]
\small
\centering
\caption{Absolute distance between the recovered and the configured HEXACO level, averaged over cases. Lower distance is better and bolded per axis.}
\label{tab:hexaco_diffs}
\begin{tabular}{llccccccc}
\toprule
Patient Simulator & Rater & H & E & X & A & C & O & Mean \\
\midrule
\multirow{3}{*}{\ours}
  & Autorater & 0.36 & 0.52 & 0.97 & 0.40 & 0.36 & 0.17 & 0.46 \\
  & Human     & 0.59 & 0.55 & 0.60 & 0.46 & 0.47 & 0.54 & 0.53 \\
  & Mean      & 0.47 & \textbf{0.53} & \textbf{0.78} & \textbf{0.43} & \textbf{0.41} & \textbf{0.36} & \textbf{0.50} \\
\midrule
\multirow{3}{*}{PatientSim}
  & Autorater & 0.15 & 1.23 & 1.40 & 0.69 & 0.47 & 0.58 & 0.75 \\
  & Human     & 0.51 & 0.87 & 0.54 & 0.85 & 0.65 & 0.80 & 0.70 \\
  & Mean      & \textbf{0.33} & 1.05 & 0.97 & 0.77 & 0.56 & 0.69 & 0.73 \\
\midrule
PatientSim $-$ \ours & Mean & $-0.14$ & $+0.51$ & $+0.18$ & $+0.34$ & $+0.14$ & $+0.34$ & $+0.23$ \\
\bottomrule
\end{tabular}
\end{table*}

\paragraph{Recovery of configured personality.}
First, we test whether the configured HEXACO traits are recoverable from a conversation. Our clinicians and an LLM-based autorater attempt to recover the configuration of all six axes, while blinded to the axis being manipulated. Table~\ref{tab:hexaco_diffs} reports how far the rated level lands from the configured one. Pooled over the six axes, \ours stays closer to its configuration than the paired PatientSim configuration under both rater types. All nine clinicians agree on the difference tendency when taken individually over $108$ ratings. Clinicians rating the same conversation agree moderately (Krippendorff's $\alpha = 0.50$; $92\%$ of rating pairs within one level). The autorater covers all $360$ distinct conversations; on the conversations both rated, it agrees closely with the clinicians on the pooled distance. Re-scoring with \claude{}, which is involved neither in generation nor in prompt optimization, gives a Spearman $\rho = 0.91$ with \gemini{} on the mean recovered level per persona and axis (Appendix~\ref{app:extreme_study}). The recovered personality is consistent across the encounter: restricting the autorater to later turns leaves the pooled distance unchanged.

\paragraph{Diversity of conversational behavior.}
To measure behavioral diversity, we generate conversations for 30 encounters under the six extreme personas plus the default, embed the patient turns of each conversation with \qwen{}, and compare them by cosine distance. Holding the clinical case fixed, the seven personas of \ours land further apart than those of PatientSim ($0.77$ against $0.72$ mean pairwise distance over the $30$ cases, $p<10^{-4}$). This holds in $25/30$ direct comparisons (Figure~\ref{fig:case_spread} in Appendix~\ref{app:extreme_study}).
Across \emph{different} cases, the \emph{same} persona has a 5-gram overlap of $0.002$ for \ours against $0.031$ for PatientSim, whose default persona, e.g., uses ``thank you for your help today'' in $27/30$ encounters. \ours varies more because its latent patient state differs in every encounter instead of following fixed prompt templates.

\subsection{Downstream Benchmark}\label{sec:results_downstream}

Many clinical problems have a non-medical cause, e.g., uncontrolled diabetes because the patient dislikes pricking their finger. A treatment plan that ignores those causes is called a \emph{contextual error}: medically correct, but bound to fail for this patient. Good patient-centered decision making requires that these factors are probed for and the care plan adapted as necessary.~\citep{weiner_contextual_2010}. The VHA dataset has expert annotations of each encounter's problems and their contextual causes (e.g., cost, transportation, health beliefs)~\citep{vha_2022}. We add each problem and its cause to the case description of $92$ encounters, identically for all simulators.

Three LLMs (\gemini, \geminiflash, \claude) then act as the doctor. Seeing only the clinician-observable cues of each problem, they interview the patient for up to $20$ turns, instructed to explore the circumstances behind every problem, and then state the cause and their plan. As reference, the same LLMs read the transcript of the real encounter, in which the cause came up, with the real physician's care-plan removed so that the plan cannot be copied. A simulator on which the task is easier than on this reference volunteers more information than real patients would do in encounters. \geminipro{} judges whether the stated cause matches the coded one and whether the plan ignores it.

\begin{figure}[t]
\centering
\includegraphics[width=\linewidth]{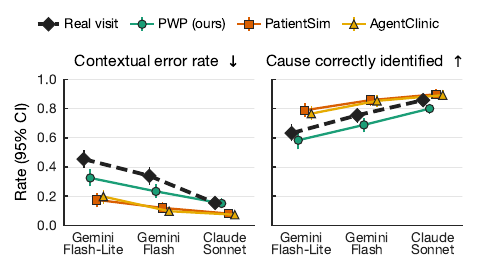}
\caption{Performance metrics of different doctor LLMs in interviewing a simulated patient (solid) or on the real encounter (dashed). Error bars show bootstrap $95\%$ CIs.}
\label{fig:downstream}
\end{figure}

As shown in Figure~\ref{fig:downstream}, PatientSim and AgentClinic make the task too easy. Every model commits fewer contextual errors than on the real encounter, so a benchmark built on them overestimates clinical performance. With \ours, the contextual error is closest to the real encounter for every model and matches it for \claude. Averaged over the three models, \ours deviates from the real error rate by $0.08$, against $0.19$ for both baselines. Cause identification is slightly harder with \ours than on the real encounter, whereas the baselines make it easier. All simulators preserve the ranking of the three models, but the baselines push every model towards the ceiling: they shrink the gap between the strongest and weakest model in cause identification from $0.23$ to $0.11$--$0.13$, while \ours keeps it at $0.22$. Extreme personas raise the difficulty further: a withholding patient ($\persH = \persC = 3$) lowers correct cause identification by $0.32$, an emotional patient with a prior belief ($\persE = \persO = 3$) by $0.15$ (cf Appendix~\ref{app:downstream}).

\section{Discussion}
We introduced \ours to address the two failure modes that limit LLM-based patient simulators: \emph{oversharing} of information the physician never asked for, and \emph{uniformity} across cooperative, knowledgeable, and ultimately indistinguishable patients. The query-conditioned disclosure grid removes unrequested fields from the prompt, preventing oversharing, and the HEXACO-parametrized latent role couples personality variation to controllable, recoverable conversational behavior. Our physician-aligned protocol enables the first quantitative head-to-head comparison of patient simulators, whose realism reference is a real patient rather than an actor. Clinicians judge \ours patients as real more often than all other patients, including the original reference. This is because the reference is a transcribed speech that is sometimes disfluent, which some clinicians read as authentic and others as broken. This is also seen in the modest inter-rater agreement of these cases. This shows that realism can be highly subjective and should thus be clearly defined based on the context of the simulator's use-case.
\ours disclosure pacing follows that of real patients, addressing the oversharing that all prior simulators exhibit, the configured HEXACO traits are reliably recoverable from conversations and are substantially more diverse than the most prominent other parametrizable baseline. In the downstream benchmark, \ours reproduces the difficulty of real patients, whereas simulators that volunteer the cause make the task easier and compress the differences between models.

\paragraph{Limitations.}
Our evaluation is grounded in real physician--patient encounters recorded during routine primary-care visits of US veterans, which strengthens validity but narrows the population and setting. Those patients tend to be older, and the majority are male. Further, all encounters come from a single health system, so our findings may not transfer directly to other specialties, settings, or populations. Realistic simulation should complement, not replace, evaluation with human participants. \ours is best positioned as a screening layer that allows for faster development cycles before costly yet important human studies are undertaken.

\iffinal
\acks{This work was funded in part by Google, and conducted with feedback and manuscript review from Google researchers. Google researchers did not access the data nor code.}
\fi

\bibliography{ref}

\begin{thebibliography}{43}
\providecommand{\natexlab}[1]{#1}
\providecommand{\url}[1]{\texttt{#1}}
\expandafter\ifx\csname urlstyle\endcsname\relax
  \providecommand{\doi}[1]{doi: #1}\else
  \providecommand{\doi}{doi: \begingroup \urlstyle{rm}\Url}\fi

\bibitem[Ashton and Lee(2007)]{ashton_empirical_2007}
Michael~C. Ashton and Kibeom Lee.
\newblock Empirical, theoretical, and practical advantages of the {HEXACO} model of personality structure.
\newblock \emph{Personality and Social Psychology Review: An Official Journal of the Society for Personality and Social Psychology, Inc}, 11\penalty0 (2):\penalty0 150--166, 2007.

\bibitem[Bean et~al.(2026)Bean, Payne, Parsons, Kirk, Ciro, Mosquera-Gómez, Hincapié~M, Ekanayaka, Tarassenko, Rocher, and Mahdi]{bean_reliability_2026}
Andrew~M. Bean, Rebecca~Elizabeth Payne, Guy Parsons, Hannah~Rose Kirk, Juan Ciro, Rafael Mosquera-Gómez, Sara Hincapié~M, Aruna~S. Ekanayaka, Lionel Tarassenko, Luc Rocher, and Adam Mahdi.
\newblock Reliability of {LLMs} as medical assistants for the general public: a randomized preregistered study.
\newblock \emph{Nature Medicine}, 32\penalty0 (2):\penalty0 609--615, 2026.

\bibitem[Cook et~al.(2025)Cook, Overgaard, Pankratz, Del~Fiol, and Aakre]{cook_virtual_2025}
David~A Cook, Joshua Overgaard, V~Shane Pankratz, Guilherme Del~Fiol, and Chris~A Aakre.
\newblock Virtual {Patients} {Using} {Large} {Language} {Models}: {Scalable}, {Contextualized} {Simulation} of {Clinician}-{Patient} {Dialogue} {With} {Feedback}.
\newblock \emph{Journal of Medical Internet Research}, 27:\penalty0 e68486, 2025.

\bibitem[Du et~al.(2025)Du, Zheng, Hu, Xu, Li, Sun, Chen, Wu, Cai, and Ying]{du_llms_2025}
Zhuoyun Du, Lujie Zheng, Renjun Hu, Yuyang Xu, Xiawei Li, Ying Sun, Wei Chen, Jian Wu, Haolei Cai, and Haochao Ying.
\newblock {LLMs} {Can} {Simulate} {Standardized} {Patients} via {Agent} {Coevolution}.
\newblock In Wanxiang Che, Joyce Nabende, Ekaterina Shutova, and Mohammad~Taher Pilehvar, editors, \emph{Proceedings of the 63rd {Annual} {Meeting} of the {Association} for {Computational} {Linguistics} ({Volume} 1: {Long} {Papers})}, pages 17278--17306, Vienna, Austria, 2025. Association for Computational Linguistics.

\bibitem[Fajardo et~al.(2025)Fajardo, Proniakin, Gruber, and Marinescu]{fajardo_medpi_2025}
Diego Fajardo, Oleksii Proniakin, Victoria-Elisabeth Gruber, and Razvan Marinescu.
\newblock {MedPI}: {Evaluating} {AI} {Systems} in {Medical} {Patient}-facing {Interactions}, 2025.
\newblock arXiv:2601.04195 [cs].

\bibitem[Fan et~al.(2025)Fan, Wei, Tang, Chen, Siyuan, Wei, and Huang]{fan_ai_2025}
Zhihao Fan, Lai Wei, Jialong Tang, Wei Chen, Wang Siyuan, Zhongyu Wei, and Fei Huang.
\newblock {AI} {Hospital}: {Benchmarking} {Large} {Language} {Models} in a {Multi}-agent {Medical} {Interaction} {Simulator}.
\newblock In Owen Rambow, Leo Wanner, Marianna Apidianaki, Hend Al-Khalifa, Barbara~Di Eugenio, and Steven Schockaert, editors, \emph{Proceedings of the 31st {International} {Conference} on {Computational} {Linguistics}}, pages 10183--10213, Abu Dhabi, UAE, 2025. Association for Computational Linguistics.

\bibitem[Feinstein and Cicchetti(1990)]{feinstein_high_1990}
Alvan~R. Feinstein and Domenic~V. Cicchetti.
\newblock High agreement but low kappa: {I}. {The} problems of two paradoxes.
\newblock \emph{Journal of Clinical Epidemiology}, 43\penalty0 (6):\penalty0 543--549, 1990.

\bibitem[Gong et~al.(2025)Gong, Bang, Lee, and Baik]{gong_knowledge-practice_2025}
Eun~Jeong Gong, Chang~Seok Bang, Jae~Jun Lee, and Gwang~Ho Baik.
\newblock Knowledge-{Practice} {Performance} {Gap} in {Clinical} {Large} {Language} {Models}: {Systematic} {Review} of 39 {Benchmarks}.
\newblock \emph{Journal of Medical Internet Research}, 27\penalty0 (1):\penalty0 e84120, 2025.

\bibitem[Gwet(2008)]{gwet_computing_2008}
Kilem~Li Gwet.
\newblock Computing inter-rater reliability and its variance in the presence of high agreement.
\newblock \emph{British Journal of Mathematical and Statistical Psychology}, 61\penalty0 (1):\penalty0 29--48, 2008.

\bibitem[Hager et~al.(2024)Hager, Jungmann, Holland, Bhagat, Hubrecht, Knauer, Vielhauer, Makowski, Braren, Kaissis, and Rueckert]{hager_evaluation_2024}
Paul Hager, Friederike Jungmann, Robbie Holland, Kunal Bhagat, Inga Hubrecht, Manuel Knauer, Jakob Vielhauer, Marcus Makowski, Rickmer Braren, Georgios Kaissis, and Daniel Rueckert.
\newblock Evaluation and mitigation of the limitations of large language models in clinical decision-making.
\newblock \emph{Nature Medicine}, 30\penalty0 (9):\penalty0 2613--2622, 2024.

\bibitem[Hayes and Krippendorff(2007)]{hayes_answering_2007}
Andrew~F. Hayes and Klaus Krippendorff.
\newblock Answering the call for a standard reliability measure for coding data.
\newblock \emph{Communication Methods and Measures}, 1\penalty0 (1):\penalty0 77--89, 2007.

\bibitem[Holderried et~al.(2024)Holderried, Stegemann-Philipps, Herschbach, Moldt, Nevins, Griewatz, Holderried, Herrmann-Werner, Festl-Wietek, and Mahling]{holderried_generative_2024}
Friederike Holderried, Christian Stegemann-Philipps, Lea Herschbach, Julia-Astrid Moldt, Andrew Nevins, Jan Griewatz, Martin Holderried, Anne Herrmann-Werner, Teresa Festl-Wietek, and Moritz Mahling.
\newblock A {Generative} {Pretrained} {Transformer} ({GPT})-{Powered} {Chatbot} as a {Simulated} {Patient} to {Practice} {History} {Taking}: {Prospective}, {Mixed} {Methods} {Study}.
\newblock \emph{JMIR medical education}, 10:\penalty0 e53961, 2024.

\bibitem[Hosseinpourkhoshkbari et~al.(2025)Hosseinpourkhoshkbari, Huang, Muttreja, and Golden]{hosseinpourkhoshkbari_spot_2025}
Reyhaneh Hosseinpourkhoshkbari, Wei-chen Huang, Suvel Muttreja, and Richard~M. Golden.
\newblock Can {You} {Spot} the {Virtual} {Patient} ({VP})? {Expert} {Evaluation}, {Turing} {Test}, {Linguistic} {Analysis}, and {Semantic} {Similarity} {Analysis}.
\newblock In \emph{NeurIPS 2025 Workshop on {GenAI} for {Health}: {Potential}, {Trust}, and {Policy} {Compliance}}, 2025.

\bibitem[Jiang et~al.(2026)Jiang, Zhang, Zhu, Bai, and Zhao]{jiang_multi-stage_2026}
Shijie Jiang, Zefan Zhang, Kehua Zhu, Tian Bai, and Ruihong Zhao.
\newblock Multi-{Stage} {Patient} {Role}-{Playing} {Framework} for {Realistic} {Clinical} {Interactions}, 2026.
\newblock arXiv:2601.10951 [cs].

\bibitem[Johri et~al.(2025)Johri, Jeong, Tran, Schlessinger, Wongvibulsin, Barnes, Zhou, Cai, Van~Allen, Kim, Daneshjou, and Rajpurkar]{johri_evaluation_2025}
Shreya Johri, Jaehwan Jeong, Benjamin~A. Tran, Daniel~I. Schlessinger, Shannon Wongvibulsin, Leandra~A. Barnes, Hong-Yu Zhou, Zhuo~Ran Cai, Eliezer~M. Van~Allen, David Kim, Roxana Daneshjou, and Pranav Rajpurkar.
\newblock An evaluation framework for clinical use of large language models in patient interaction tasks.
\newblock \emph{Nature Medicine}, 31\penalty0 (1):\penalty0 77--86, 2025.

\bibitem[Kim and Yoon(2025)]{kim_questioning_2025}
Siun Kim and Hyung-Jin Yoon.
\newblock Questioning {Our} {Questions}: {How} {Well} {Do} {Medical} {QA} {Benchmarks} {Evaluate} {Clinical} {Capabilities} of {Language} {Models}?
\newblock In Dina Demner-Fushman, Sophia Ananiadou, Makoto Miwa, and Junichi Tsujii, editors, \emph{Proceedings of the 24th {Workshop} on {Biomedical} {Language} {Processing}}, pages 274--296, Vienna, Austria, 2025. Association for Computational Linguistics.

\bibitem[Kincaid et~al.(1975)Kincaid, Fishburne, Rogers, and Chissom]{kincaid_derivation_1975}
J.~Peter Kincaid, Robert~P. Fishburne, {Jr.}, Richard~L. Rogers, and Brad~S. Chissom.
\newblock Derivation of {New} {Readability} {Formulas} ({Automated} {Readability} {Index}, {Fog} {Count} and {Flesch} {Reading} {Ease} {Formula}) for {Navy} {Enlisted} {Personnel}:.
\newblock Technical report, Defense Technical Information Center, Fort Belvoir, VA, 1975.

\bibitem[Kyung et~al.(2025)Kyung, Chung, Bae, Kim, Sohn, Kim, Kim, and Choi]{kyung_patientsim_2025}
Daeun Kyung, Hyunseung Chung, Seongsu Bae, Jiho Kim, Jae~Ho Sohn, Taerim Kim, Soo~Kyung Kim, and Edward Choi.
\newblock {PatientSim}: {A} {Persona}-{Driven} {Simulator} for {Realistic} {Doctor}-{Patient} {Interactions}.
\newblock In \emph{The Thirty-ninth Annual Conference on Neural Information Processing Systems Datasets and Benchmarks Track}, 2025.

\bibitem[Liao et~al.(2024)Liao, Meng, Wang, Liu, Wang, and Wang]{liao_automatic_2024}
Yusheng Liao, Yutong Meng, Yuhao Wang, Hongcheng Liu, Yanfeng Wang, and Yu~Wang.
\newblock Automatic {Interactive} {Evaluation} for {Large} {Language} {Models} with {State} {Aware} {Patient} {Simulator}, 2024.
\newblock arXiv:2403.08495 [cs.CL].

\bibitem[Li\'{e}vin et~al.(2026)Li\'{e}vin, Schmidgall, Strother, Bijamov, Goel, Palepu, Park, Balazadeh, Sun, Guerard, Chen, Steiner, Dhillon, Azar, Mehta, Spetsieris, Shah, Abdelrahim, Dahiya, Liu, Chou, Matias, Hassidim, Webster, Le, Hadsell, Barral, Radebaugh, Faust, Azizi, Schaekermann, Chen, Tu, Racz, and Yang]{lievin_residencyrl_2026}
Valentin Li\'{e}vin, Samuel Schmidgall, Tim Strother, Alex Bijamov, Akshay Goel, Anil Palepu, Chunjong Park, Vahid Balazadeh, Min~Woo Sun, Marius Guerard, Justin Chen, Dave Steiner, Vikram Dhillon, Ibrahim Azar, Akhil Mehta, Nicholas Spetsieris, Shilpan Shah, Maen Abdelrahim, Amit Dahiya, Yun Liu, Katherine Chou, Yossi Matias, Avinatan Hassidim, Dale~R. Webster, Quoc~V. Le, Raia Hadsell, Joelle Barral, Carey Radebaugh, Aleksandra Faust, Shekoofeh Azizi, Mike Schaekermann, Po-Hsuan~Cameron Chen, Tao Tu, David Racz, and Lin Yang.
\newblock {ResidencyRL}: {Reinforcement} {Learning} in {Simulated} {Clinical} {Environments}, 2026.
\newblock arXiv:2608.07418 [cs].

\bibitem[Lu et~al.(2022)Lu, Bartolo, Moore, Riedel, and Stenetorp]{lu_fantastically_2022}
Yao Lu, Max Bartolo, Alastair Moore, Sebastian Riedel, and Pontus Stenetorp.
\newblock Fantastically {Ordered} {Prompts} and {Where} to {Find} {Them}: {Overcoming} {Few}-{Shot} {Prompt} {Order} {Sensitivity}.
\newblock In Smaranda Muresan, Preslav Nakov, and Aline Villavicencio, editors, \emph{Proceedings of the 60th {Annual} {Meeting} of the {Association} for {Computational} {Linguistics} ({Volume} 1: {Long} {Papers})}, pages 8086--8098, Dublin, Ireland, 2022. Association for Computational Linguistics.

\bibitem[Mansouri et~al.(2026)Mansouri, Marvania, Shihorkar, Tran, Shafiei, Fazli, Li, and Zhu]{mansouri_verisim_2026}
Sina Mansouri, Mohit Marvania, Vibhavari~Ashok Shihorkar, Han~Ngoc Tran, Kazhal Shafiei, Mehrdad Fazli, Yikuan Li, and Ziwei Zhu.
\newblock {VeriSim}: {A} {Configurable} {Framework} for {Evaluating} {Medical} {AI} {Under} {Realistic} {Patient} {Noise}, 2026.
\newblock arXiv:2604.10441 [cs].

\bibitem[Matos et~al.(2026)Matos, Buege, Cheung, Collins, Dhiman, Li, Mao, Nelson, Ouroutzoglou, Varghese, and Amar]{matos_complexities_2026}
Jo\~{a}o Matos, Olivia Buege, Donny Cheung, Gary~S. Collins, Paula Dhiman, Nan Li, Bingyu Mao, Benjamin~W. Nelson, Michail Ouroutzoglou, Paul Varghese, and Jonathan Amar.
\newblock The complexities of patient-centred conversational artificial intelligence, 2026.
\newblock arXiv:2607.08625 [cs].

\bibitem[Pletzer et~al.(2024)Pletzer, Thielmann, and Zettler]{pletzer_who_2024}
Jan~Luca Pletzer, Isabel Thielmann, and Ingo Zettler.
\newblock Who is healthier? {A} meta-analysis of the relations between the {HEXACO} personality domains and health outcomes.
\newblock \emph{European Journal of Personality}, 38\penalty0 (2):\penalty0 342--364, 2024.

\bibitem[Roccas et~al.(2002)Roccas, Sagiv, Schwartz, and Knafo]{roccas_big_2002}
Sonia Roccas, Lilach Sagiv, Shalom~H. Schwartz, and Ariel Knafo.
\newblock The {Big} {Five} {Personality} {Factors} and {Personal} {Values}.
\newblock \emph{Personality and Social Psychology Bulletin}, 28\penalty0 (6):\penalty0 789--801, 2002.

\bibitem[Sadjoli et~al.(2025)Sadjoli, Siefken, Ghosh, Mai, and Dahlmeier]{sadjoli_optimization_2025}
Nicholas Sadjoli, Tim Siefken, Atin Ghosh, Yifan Mai, and Daniel Dahlmeier.
\newblock Optimization before {Evaluation}: {Evaluation} with {Unoptimized} {Prompts} {Can} be {Misleading}.
\newblock In Georg Rehm and Yunyao Li, editors, \emph{Proceedings of the 63rd {Annual} {Meeting} of the {Association} for {Computational} {Linguistics} ({Volume} 6: {Industry} {Track})}, pages 619--638, Vienna, Austria, 2025. Association for Computational Linguistics.

\bibitem[Schmidgall et~al.(2026)Schmidgall, Ziaei, Harris, Kim, Reis, Jopling, and Moor]{schmidgall_agentclinic_2026}
Samuel Schmidgall, Rojin Ziaei, Carl Harris, Ji~Woong Kim, Eduardo~Pontes Reis, Jeffrey Jopling, and Michael Moor.
\newblock {AgentClinic}: a multimodal benchmark for tool-using clinical {AI} agents.
\newblock \emph{npj Digital Medicine}, 2026.

\bibitem[Singhal et~al.(2023)Singhal, Azizi, Tu, Mahdavi, Wei, Chung, Scales, Tanwani, Cole-Lewis, Pfohl, Payne, Seneviratne, Gamble, Kelly, Babiker, Schärli, Chowdhery, Mansfield, Demner-Fushman, Agüera~y Arcas, Webster, Corrado, Matias, Chou, Gottweis, Tomasev, Liu, Rajkomar, Barral, Semturs, Karthikesalingam, and Natarajan]{singhal_large_2023}
Karan Singhal, Shekoofeh Azizi, Tao Tu, S.~Sara Mahdavi, Jason Wei, Hyung~Won Chung, Nathan Scales, Ajay Tanwani, Heather Cole-Lewis, Stephen Pfohl, Perry Payne, Martin Seneviratne, Paul Gamble, Chris Kelly, Abubakr Babiker, Nathanael Schärli, Aakanksha Chowdhery, Philip Mansfield, Dina Demner-Fushman, Blaise Agüera~y Arcas, Dale Webster, Greg~S. Corrado, Yossi Matias, Katherine Chou, Juraj Gottweis, Nenad Tomasev, Yun Liu, Alvin Rajkomar, Joelle Barral, Christopher Semturs, Alan Karthikesalingam, and Vivek Natarajan.
\newblock Large language models encode clinical knowledge.
\newblock \emph{Nature}, 620\penalty0 (7972):\penalty0 172--180, 2023.

\bibitem[Srirag et~al.(2026)Srirag, Nguyen, Joshi, Narasimhan, and Kanhere]{srirag_triagesim_2026}
Dipankar Srirag, Quoc~Dung Nguyen, Aditya Joshi, Padmanesan Narasimhan, and Salil Kanhere.
\newblock {TriageSim}: {A} {Conversational} {Emergency} {Triage} {Simulation} {Framework} from {Structured} {Electronic} {Health} {Records}, 2026.
\newblock arXiv:2603.10035 [cs] version: 1.

\bibitem[Tang et~al.(2023)Tang, Sun, Idnay, Nestor, Soroush, Elias, Xu, Ding, Durrett, Rousseau, Weng, and Peng]{tang_evaluating_2023}
Liyan Tang, Zhaoyi Sun, Betina Idnay, Jordan~G. Nestor, Ali Soroush, Pierre~A. Elias, Ziyang Xu, Ying Ding, Greg Durrett, Justin~F. Rousseau, Chunhua Weng, and Yifan Peng.
\newblock Evaluating large language models on medical evidence summarization.
\newblock \emph{npj Digital Medicine}, 6\penalty0 (1):\penalty0 158, 2023.

\bibitem[{U.S. Department of Veterans Affairs}(2022)]{vha_2022}
{U.S. Department of Veterans Affairs}.
\newblock Physician-patient transcripts with {4C} coding analysis from the {Contextualizing Care} research program.
\newblock Dataset, ID \texttt{4qbs-wgct}, 2022.
\newblock URL \url{https://www.data.va.gov/dataset/Physician-patient-transcripts-with-4C-coding-analy/4qbs-wgct}.
\newblock 405 de-identified primary care encounter transcripts with Content Coding for Contextualization of Care (4C) annotations. Licensed CC0 1.0. Accessed 2026-08-26.

\bibitem[Vatanparvar et~al.(2026)Vatanparvar, Joshi, Xenochristou, Hashemi, Kasu, Bansal, Lopez-Martinez, Nema, Ganesan, Kimbrough, Woody, Rao, Hakkani-Tur, and Schulz-Mahlendorf]{vatanparvar_patientagentbench_2026}
Korosh Vatanparvar, Ashutosh Joshi, Maria Xenochristou, Mohammad~Abuzar Hashemi, Prasad Kasu, Deepak Bansal, Daniel Lopez-Martinez, Anchal Nema, Ramya Ganesan, Will Kimbrough, Alex Woody, Yadunandana Rao, Dilek Hakkani-Tur, and Wilko Schulz-Mahlendorf.
\newblock {PatientAgentBench}: {A} {Benchmark} {Framework} for {Evaluating} {Patient}-{Facing} {Health} {AI} {Agents}, 2026.
\newblock arXiv:2607.25485 [cs].

\bibitem[Wang et~al.(2025)Wang, Li, Chen, Yuan, Yang, and Wong]{wang_exploring_2025}
Shuo Wang, Renhao Li, Xi~Chen, Yulin Yuan, Min Yang, and Derek~F. Wong.
\newblock Exploring the {Impact} of {Personality} {Traits} on {LLM} {Bias} and {Toxicity}.
\newblock In Christos Christodoulopoulos, Tanmoy Chakraborty, Carolyn Rose, and Violet Peng, editors, \emph{Proceedings of the 2025 {Conference} on {Empirical} {Methods} in {Natural} {Language} {Processing}}, pages 4125--4143, Suzhou, China, 2025. Association for Computational Linguistics.

\bibitem[Weiner et~al.(2010)Weiner, Schwartz, Weaver, Goldberg, Yudkowsky, Sharma, Binns-Calvey, Preyss, Schapira, Persell, Jacobs, and Abrams]{weiner_contextual_2010}
Saul~J. Weiner, Alan Schwartz, Frances Weaver, Julie Goldberg, Rachel Yudkowsky, Gunjan Sharma, Amy Binns-Calvey, Ben Preyss, Marilyn~M. Schapira, Stephen~D. Persell, Elizabeth Jacobs, and Richard~I. Abrams.
\newblock Contextual errors and failures in individualizing patient care: a multicenter study.
\newblock \emph{Annals of Internal Medicine}, 153\penalty0 (2):\penalty0 69--75, 2010.

\bibitem[Weiner et~al.(2013)Weiner, Schwartz, Sharma, Binns-Calvey, Ashley, Kelly, Dayal, Patel, Weaver, and Harris]{weiner_patient-centered_2013}
Saul~J. Weiner, Alan Schwartz, Gunjan Sharma, Amy Binns-Calvey, Naomi Ashley, Brendan Kelly, Amit Dayal, Sonal Patel, Frances~M. Weaver, and Ilene Harris.
\newblock Patient-centered decision making and health care outcomes: an observational study.
\newblock \emph{Annals of Internal Medicine}, 158\penalty0 (8):\penalty0 573--579, 2013.

\bibitem[Wu et~al.(2026{\natexlab{a}})Wu, Wu, Hom, Yi, Zhang, Lozano, Nirschl, Tangney, Byram, Dymm, Annapureddy, Topol, Ouyang, and Zou]{wu_medarena_2026}
Eric Wu, Kevin Wu, Jason Hom, Paul~H. Yi, Angela Zhang, Alejandro Lozano, Jeff Nirschl, Jeff Tangney, Kevin Byram, Braydon Dymm, Narender Annapureddy, Eric Topol, David Ouyang, and James Zou.
\newblock {MedArena}: {Comparing} {LLMs} for {Medicine}-in-the-{Wild} {Clinician} {Preferences}, 2026{\natexlab{a}}.
\newblock arXiv:2603.15677 [cs].

\bibitem[Wu et~al.(2026{\natexlab{b}})Wu, Han, Zhang, Li, Jiang, Lu, Zhang, Xu, Ming, Wang, and Wen]{wu_simulation_2026}
Ping Wu, Yu~Han, Jing Zhang, Yunqi Li, Mengna Jiang, Xinyu Lu, Haibin Zhang, Danyang Xu, Hao Ming, Lihong Wang, and Qingping Wen.
\newblock From simulation to pedagogy: structured {AI} standardized patients for clinical communication training validated through multi-model and randomized evaluation, 2026{\natexlab{b}}.
\newblock medRxiv 2026.04.26.26351793.

\bibitem[Yim et~al.(2023)Yim, Fu, Ben~Abacha, Snider, Lin, and Yetisgen]{yim_aci-bench_2023}
Wen-wai Yim, Yujuan Fu, Asma Ben~Abacha, Neal Snider, Thomas Lin, and Meliha Yetisgen.
\newblock Aci-bench: a {Novel} {Ambient} {Clinical} {Intelligence} {Dataset} for {Benchmarking} {Automatic} {Visit} {Note} {Generation}.
\newblock \emph{Scientific Data}, 10\penalty0 (1):\penalty0 586, 2023.

\bibitem[Zehle et~al.(2025)Zehle, Schlager, Heiß, and Feurer]{zehle_capo_2025}
Tom Zehle, Moritz Schlager, Timo Heiß, and Matthias Feurer.
\newblock {CAPO}: {Cost}-{Aware} {Prompt} {Optimization}.
\newblock In \emph{Proceedings of the {Fourth} {International} {Conference} on {Automated} {Machine} {Learning}}, pages 18/1--45. PMLR, 2025.

\bibitem[Zehle et~al.(2026)Zehle, Hei{\ss}, Schlager, A{\ss}enmacher, and Feurer]{zehle_promptolution_2026}
Tom Zehle, Timo Hei{\ss}, Moritz Schlager, Matthias A{\ss}enmacher, and Matthias Feurer.
\newblock promptolution: A unified, modular framework for prompt optimization.
\newblock In \emph{Proceedings of the 19th Conference of the {E}uropean Chapter of the {A}ssociation for {C}omputational {L}inguistics (Volume 3: System Demonstrations)}, pages 282--296, Rabat, Morocco, March 2026. Association for Computational Linguistics.

\bibitem[Zeng et~al.(2007)Zeng, Tse, Divita, Keselman, Crowell, Browne, Goryachev, and Ngo]{zeng_term_2007}
Qing Zeng, Tony Tse, Guy Divita, Alla Keselman, Jonathan Crowell, Allen Browne, Sergey Goryachev, and Long Ngo.
\newblock Term {Identification} {Methods} for {Consumer} {Health} {Vocabulary} {Development}.
\newblock \emph{Journal of Medical Internet Research}, 9\penalty0 (1):\penalty0 e606, 2007.

\bibitem[Zhang et~al.(2026)Zhang, Liu, Wang, Zhou, Xie, and Wang]{zhang_human_2026}
Bingquan Zhang, Xiaoxiao Liu, Yuchi Wang, Lei Zhou, Qianqian Xie, and Benyou Wang.
\newblock Human or {LLM} as {Standardized} {Patients}? {A} {Comparative} {Study} for {Medical} {Education}, 2026.
\newblock arXiv:2511.14783 [cs.CL].

\bibitem[Zhou et~al.(2023)Zhou, Muresanu, Han, Paster, Pitis, Chan, and Ba]{zhou_large_2022}
Yongchao Zhou, Andrei~Ioan Muresanu, Ziwen Han, Keiran Paster, Silviu Pitis, Harris Chan, and Jimmy Ba.
\newblock Large {Language} {Models} are {Human}-{Level} {Prompt} {Engineers}.
\newblock In \emph{The Eleventh International Conference on Learning Representations}, 2023.

\end{thebibliography}

\newpage
\appendix
\onecolumn
\raggedbottom

\section{Structured Case Description Fields}\label{app:case_fields}

Table~\ref{tab:case_fields} enumerates the named fields of the structured case description $\case$ used by \textit{\ours}, together with their assignment to the three semantic subgroups $\Fper$, $\Flife$, and $\Fmed$. Personal fields are always visible to the conversational model. Lifestyle fields are subject to an honesty-humility-dependent disclosure mask with three variants (original, downplayed, denied). Medical fields are subject to a conscientiousness-dependent recall mask (original, fuzzy, omitted) that is relaxed when the corresponding content is explicitly elicited by the physician.

\begin{table*}[h]
\centering
\small
\caption{Named fields of the structured case description and their assignment to the subgroups $\Fper$ (personal background), $\Flife$ (sensitive lifestyle habits), and $\Fmed$ (clinical facts). Field names match the keys used in the reference implementation.}
\label{tab:case_fields}
\begin{tabular}{ll}
\toprule
\textbf{Field} & \textbf{Content} \\
\midrule
\multicolumn{2}{l}{\textit{$\Fper$ — Personal background (always visible)}} \\
\midrule
\texttt{age} & Patient age in years. \\
\texttt{gender} & Reported gender. \\
\texttt{marital\_status} & Relationship status (e.g., single, married, widowed). \\
\texttt{children} & Number and/or presence of children. \\
\texttt{living\_situation} & Housing and household composition. \\
\texttt{occupation} & Current or most recent occupation. \\
\texttt{insurance} & Insurance status or provider type. \\
\texttt{arrival\_transport} & Mode of arrival at the encounter (e.g., walk-in, ambulance). \\
\texttt{chiefcomplaint} & Primary complaint stated on presentation. \\
\midrule
\multicolumn{2}{l}{\textit{$\Flife$ — Sensitive lifestyle habits (honesty-humility-gated disclosure)}} \\
\midrule
\texttt{tobacco} & Tobacco use and frequency. \\
\texttt{alcohol} & Alcohol consumption pattern. \\
\texttt{illicit\_drug} & Use of non-prescribed or illicit drugs. \\
\texttt{sexual\_history} & Relevant sexual history. \\
\texttt{exercise} & Physical activity routine. \\
\midrule
\multicolumn{2}{l}{\textit{$\Fmed$ — Clinical facts (conscientiousness-gated recall)}} \\
\midrule
\texttt{allergies} & Known allergies and associated reactions. \\
\texttt{family\_medical\_history} & Relevant medical history in first-degree relatives. \\
\texttt{medical\_device} & Implants or assistive medical devices. \\
\texttt{medical\_history} & Past diagnoses, conditions, and procedures. \\
\texttt{present\_illness\_positive} & Affirmed symptoms of the current illness. \\
\texttt{present\_illness\_negative} & Explicitly denied symptoms (pertinent negatives). \\
\texttt{pain} & Pain location, character, and severity. \\
\texttt{medication} & Current medications and dosages. \\
\bottomrule
\end{tabular}
\end{table*}

\newpage
\section{Prompt Templates}\label{app:prompts}

This section reproduces the prompt templates used by \textit{\ours} at each stage of the simulation loop introduced in Section~\ref{sec:framework} and throughout the evaluation. Placeholders substituted at call time are shown in \textcolor{purple}{\texttt{\textless{}purple\textgreater{}}}; all other content is rendered verbatim. The prompts used for evaluation, case extraction, the autorater, and the fact-fidelity checks, are given in the reference implementation.

\begin{prompt}[breakable=true]{Latent role construction (\texttt{LATENT\_ROLE})}{latent-role}
Craft a concise, clinically precise latent role specification for a patient simulator, emphasizing only empirically observable behavioral and communicative traits that shape real-time interaction dynamics during anamnesis. The specification must synthesize the following behavioral axes into a functional, simulation-ready template while preserving the exact HEXACO personality dimensions and personal details provided---without deviation or embellishment of factual data:

\smallskip
\noindent\textbf{Key Interactional Parameters:}
\begin{enumerate}\setlength\itemsep{0pt}
  \item \textbf{Speech Clarity} -- Assessable through linguistic fluency, hesitation patterns, and syntactic precision (e.g., fragmented phrasing, circumlocution).
  \item \textbf{Information Sharing} -- Degree of openness in divulging personal history, granularity of responses, and comfort with self-exposure.
  \item \textbf{Affective Expression} -- Spectrum of emotional display (e.g., muted, exaggerated, or detached affect).
  \item \textbf{Cognitive Engagement} -- Logical consistency of replies, tangentiality, thematic alignment with clinician queries, and adaptability to probing.
  \item \textbf{Dialogue Timing} -- Response latency, contextual relevance, and conversational pacing.
\end{enumerate}

\smallskip
\noindent\textbf{Directives:}
\begin{itemize}\setlength\itemsep{0pt}
  \item Generate a stripped-down, machine-actionable internal state model, excluding extraneous narrative or speculative content.
  \item Encase the role strictly within \texttt{<role>} and \texttt{</role>} tags for seamless integration into multi-turn diagnostic simulations.
  \item Retain all provided HEXACO traits and personal details \emph{literally}---no modifications, additions, or fabrications.
\end{itemize}

\smallskip
\noindent\textbf{Deliverable:} Output solely the formatted role card, ensuring it operationalizes the above parameters into a deployable simulation framework.

\medskip
\textbf{Personal Information:} \ph{personal\_information} \\
\textbf{HEXACO Personality:} \ph{hexaco\_personality}
\end{prompt}

\begin{table*}[tb]
\centering
\caption{HEXACO level descriptors assembled into \ph{hexaco\_personality} for the latent-role construction prompt. The state variable \ph{state} is generated by prompt \texttt{E\_META}, \ph{prior\_belief} is generated by prompt \texttt{O\_META}.}
\footnotesize
\begin{tabular}{@{}p{3.3cm}p{1cm}p{\dimexpr\textwidth-3.3cm-1cm-4\tabcolsep\relax}@{}}
\toprule
\textbf{Axis} & \textbf{Level} & \textbf{Descriptor} \\
\midrule
Honesty-Humility (\texttt{H}) & 1 & You are transparent and honest, sharing all habits including substance use or non-compliance openly. \\
 & 2 & You are generally truthful but hesitant to share embarrassing habits and tend to downplay them. \\
 & 3 & You are actively hiding specific habits and may provide false information about your medical history. \\
\midrule
Emotionality (\texttt{E}) & 1 & You are emotionally detached and calm. Report facts objectively without expressing fear or distress. \\
 & 2 & You are concerned about your health. \\
 & 3 & \ph{state} \\
\midrule
eXtraversion (\texttt{X}) & 1 & You are passive and reserved. Only answer exactly what is asked using as few words as possible. \\
 & 2 & You have a natural conversational flow. Provide concise but complete answers that stay strictly on topic. \\
 & 3 & You are talkative and inquisitive. Include personal anecdotes and ask the doctor questions about the process. \\
\midrule
Agreeableness (\texttt{A}) & 1 & You trust the medical staff and the process. Be helpful and cooperative. \\
 & 2 & You are guarded and suspicious. Do not volunteer information unless the doctor asks a very specific question. \\
 & 3 & You are frustrated and skeptical of the doctor. You are easily annoyed and may bring up past bad experiences. \\
\midrule
Conscientiousness (\texttt{C}) & 1 & You remember dates, times, and dosages for all medical events. \\
 & 2 & Your memory for dates and dosages is a bit fuzzy. \\
 & 3 & You struggle to remember when symptoms started or what medications you take. \\
\midrule
Openness (\texttt{O}) & 1 & You appreciate scientific explanations and are open to any logical medical advice provided. \\
 & 2 & You are hesitant about new technology. You prefer standard, tried-and-true treatments or home remedies. \\
 & 3 & You are convinced you already know what is wrong. Based on your own research, you assume that you have: \ph{prior\_belief}. Dismiss other ideas. \\
\bottomrule
\end{tabular}
\label{tab:prompts-hexaco}
\end{table*}

\begin{prompt}[breakable=false]{Conversational system prompt (\texttt{SYS})}{conv-sys}
\ph{latent\_role}

\textbf{Patient Profile:} \ph{structured\_information}

\ph{style\_directives}

Provide your answer to the doctor's inquiry between the tags \texttt{<response>} and \texttt{</response>}. Do not include any other text in your answer.
\end{prompt}

\newpage

\begin{table*}[tb]
\centering
\caption{Conversation type descriptors selected by $\convtype$ and form descriptors selected by $\turnform_t$, assembled into \ph{style\_directives} for the conversational system prompt (Prompt~\ref{prompt:conv-sys}). The type descriptor is always included; the form descriptor is included with probability $\answerformrate$.}
\footnotesize
\begin{tabular}{@{}p{2.6cm}p{\dimexpr\textwidth-2.6cm-2\tabcolsep\relax}@{}}
\toprule
\textbf{Value} & \textbf{Descriptor} \\
\midrule
$\convtype=$ written & You are typing to the doctor in a chat window, not speaking to them. Write in complete, punctuated sentences. You compose a message before sending it, so nothing you send carries hesitation, fillers, false starts or self-corrections. \\
\midrule
$\convtype=$ spoken & You are speaking out loud, not writing. Sentences may be incomplete. Use contractions everywhere a speaker would (``I'm'', ``don't'', ``it's''). Drop the subject or auxiliary when a speaker would (``Been alright.'' not ``It has been alright.''). Never produce a tidy, well-formed written sentence when a fragment would do. In a longer answer you may hesitate, use a filler (``uh'', ``well'', ``you know''), or correct yourself mid-sentence -- occasionally, not in every answer, and never in a one-word one. \\
\midrule
$\turnform_t=$ polar & The doctor asked something answerable with yes or no. Answer with just that. Stop there. Nothing follows the yes or no -- no explanation, no qualifier, no second sentence. \\
\midrule
$\turnform_t=$ statement & The doctor did not ask you anything -- this is an acknowledgement, an explanation or small talk. A brief acknowledgement is all this needs. Do not volunteer anything. \\
\midrule
$\turnform_t=$ open & The doctor asked something that needs a real answer. Answer it in your own words. \\
\bottomrule
\end{tabular}
\label{tab:prompts-convtype}
\end{table*}

\begin{prompt}[breakable=false]{Emotional-state generator (\texttt{E\_META})}{e-meta}
Generate an emotional state of distress for this personal profile: \ph{personal\_information}

\smallskip
\noindent Requirements:
\begin{itemize}\setlength\itemsep{0pt}
  \item The state must directly affect communication style, not factual content.
  \item Specify how the distress alters speech (e.g., avoidance, repetition, irritability, catastrophizing).
  \item Keep the description short (1--2 sentences).
\end{itemize}

Output the emotional state description between the tags \texttt{<state>} and \texttt{</state>}.
\end{prompt}

\begin{prompt}[breakable=false]{Prior-belief generator (\texttt{O\_META})}{o-meta}
Based on the complaint ``\ph{chiefcomplaint}'', generate a plausible but specific self-diagnosis that a non-medical person might find on an internet forum or from a `friend.' This diagnosis should be something the patient can fixate on dogmatically. Return the self-diagnosis between the tags \texttt{<prior\_belief>} and \texttt{</prior\_belief>}.
\end{prompt}

\begin{prompt}[breakable=false]{Lifestyle downplay generator (\texttt{H\_DOWNPLAY})}{h-downplay}
You are a linguistic assistant that converts medical data into casual, minimized euphemisms. Follow the format of the examples below exactly. Provide only the tag and the content.

\smallskip
\textbf{Input:} `drinks 6 beers a day' \\
\textbf{Output:} \texttt{<phrase>}drinks a few beers\texttt{</phrase>}

\smallskip
\textbf{Input:} `Marijuana, smoked, every weekend' \\
\textbf{Output:} \texttt{<phrase>}uses a little weed sometimes\texttt{</phrase>}

\smallskip
\textbf{Input:} `\ph{lifestyle\_info}' \\
\textbf{Output:}
\end{prompt}

\begin{prompt}[breakable=false]{Medical fuzzifier (\texttt{C\_FUZZY})}{c-fuzzy}
You are a linguistic assistant that converts medical data into a fuzzy, broader description. Transform the actual content while keeping the same general meaning. Provide only the tag and the content.

\smallskip
\textbf{Input:} family\_medical\_history: `Mother had breast cancer at age 45' \\
\textbf{Output:} \texttt{<phrase>}mother had cancer, not sure what kind or when exactly\texttt{</phrase>}

\smallskip
\textbf{Input:} pain: `7--8' \\
\textbf{Output:} \texttt{<phrase>}pretty strong pain\texttt{</phrase>}

\smallskip
\textbf{Input:} \ph{medical\_info} \\
\textbf{Output:}
\end{prompt}

\begin{prompt}[breakable=false]{Relevant-field and form classifier (\texttt{CLASS\_FORM})}{class-form}
You are analyzing a medical consultation conversation to identify which fields from the patient's case description are required to answer the doctor's question.

\smallskip
\textbf{Available case description fields:} \ph{available\_fields} \\
\textbf{Doctor's Question:} \ph{question}

\smallskip
\textbf{Task:} Identify case description fields that contain information asked about in the doctor's question. Return only fields that are explicitly mentioned or clearly implied by the question.

Return a list of relevant field names from the available fields. Return an empty list if the question is too vague or no fields are directly relevant.

\smallskip
\noindent Also classify the form of the doctor's turn:
\begin{itemize}\setlength\itemsep{0pt}
  \item \texttt{polar}: a question answerable with yes or no
  \item \texttt{open}: a question that needs more than yes or no
  \item \texttt{statement}: not a question at all (an acknowledgement, an explanation, or small talk)
\end{itemize}
\end{prompt}

\begin{prompt}[breakable=false]{Doctor-utterance alignment (\texttt{PWP\_DOC\_ALIGN})}{doc-align}
\textbf{Patient message:} \ph{patient\_last\_message}

\textbf{Original doctor response:} \ph{response}

\smallskip
\textbf{Task:} Edit the doctor response so it is consistent with the patient message.

\smallskip
\noindent\textbf{Constraints:}
\begin{itemize}\setlength\itemsep{0pt}
  \item Make the smallest possible changes (minimal edits).
  \item Keep wording, structure, and length as close as possible to the original.
  \item Only modify parts that are inconsistent or fail to address the patient message.
  \item If the patient asks a question, answer it briefly within the existing response.
  \item Do not add new information or expand the response.
\end{itemize}

Return only the revised doctor response.
\end{prompt}

\begin{prompt}[breakable=false]{Human rephrase (\texttt{PWP\_REPHRASE})}{rephrase}
\textbf{System prompt:}

You are a helpful assistant that rephrases patient responses. Maintain the same meaning, information content, and tone.

\smallskip
\hrule
\smallskip

\textbf{User prompt:}

Rephrase the following patient response. Keep the response length similar to the original response:

\smallskip
\ph{real\_response}

\smallskip
Provide your answer to the doctor's inquiry between the tags \texttt{<response>} and \texttt{</response>}. Do not include any other text in your answer.
\end{prompt}

\section{Datasets} \label{app:datasets}
\paragraph{VHA.}
The VHA dataset~\citep{vha_2022} provides 405 de-identified primary-care encounter transcripts between physicians and real patients, released by the U.S. Department of Veterans Affairs as part of the Contextualizing Care research program. It consists of raw speech-to-text transcripts only, without accompanying structured patient profiles. The encounters were recorded with $139$ physicians at two hospitals~\citep{weiner_patient-centered_2013}, so the fixed physician track of the physician-aligned protocol spans a broad range of physician behavior.

\paragraph{Transcript preparation.}
Raw transcripts carry an encounter metadata header and footer, redaction masks, timestamps, and stage directions, and label speakers as \texttt{DOCTOR}/\texttt{PATIENT} alongside companions such as \texttt{WIFE} or \texttt{SECOND PERSON}. We drop everything before the first speaker label, strip metadata, masks, timestamps, and non-verbal cues, and fold every non-clinician speaker onto the patient side to obtain a two-party transcript. Encounters in which a second physician enters are discarded, since a second clinician breaks the two-party assumption underlying the physician-aligned protocol; this retains $342$ of the $405$ transcripts.

\paragraph{Case description extraction.}
Because no structured profile is provided, we derive the case description $\case$ for each retained encounter directly from its cleaned transcript with a single extraction pass of \gemini{}, prompted to fill the fixed field schema of Appendix~\ref{app:case_fields} and to emit \texttt{Unknown} for any field the encounter never mentions. Patient age is the single exception: where the encounter never states it ($25$ of the $30$ evaluation encounters), we estimate it in a separate \gemini{} pass from the life-stage cues of the transcript and write it into $\case$. No other field is inferred beyond what the transcript states, and no manual correction is applied, so the resulting $\case$ is a faithful, if incomplete, record of what the real encounter revealed. The transcript itself provides the fixed clinician track used in the physician-aligned protocol.

\paragraph{Case selection.}
A held-out split of 30 encounters forms the meta-prompt tuning corpus (Appendix~\ref{app:metaprompt}) and is excluded from every study reported in the paper. From the remainder we keep encounters with at most $100$ patient turns (longer visits exceed the context budget of the aligned protocol) and with at least $8$ of the $13$ scorable $\Flife \cup \Fmed$ fields known, so that disclosure behavior can be measured at all. This yields the $30$ encounters used in the main evaluation:

\begin{quote}\ttfamily\small\noindent
A343, B008, B273, B311, B552, D009, D289, E679, E862, F609, G360, I984, K118, K397, K792, L144, L314, M483, N112, N927, N965, P355, P441, P493, Q997, T348, T505, X294, Y477, Y768
\end{quote}

\noindent
All $30$ are run under every patient type and configuration. The clinician study draws two disjoint subsets from them: $20$ encounters for the realism task (all of the above except \texttt{L144}, \texttt{N112}, \texttt{X294} and the seven personality cases below) and a rotation pool of $7$ for the personality task (\texttt{N965}, \texttt{P355}, \texttt{P493}, \texttt{Q997}, \texttt{T348}, \texttt{Y477}, \texttt{Y768}). Three encounters are excluded from the clinician study: \texttt{L144} and \texttt{N112}, where all simulators get stuck inside a farewell loop, and \texttt{X294}, whose transcript elicits a degenerate response.

\paragraph{Domain term lexicon.}
The domain term count reported in Section~\ref{sec:results_default} is measured against the Open Access Consumer Health Vocabulary~\citep{zeng_term_2007}, a UMLS-linked lexicon of roughly $158$k consumer and professional health expressions. We match the lowercased, punctuation-stripped tokens of an utterance against its term strings and count the distinct terms found.

\section{Meta-Prompt Tuning}\label{app:metaprompt}

The latent-role construction prompt is the main persona-relevant conditioning signal that $\Fconv$ receives on every subsequent turn. Therefore, variation in its output compounds over the entire dialogue and can eventually mischaracterize individual HEXACO axes, distort personal-information exposure, or inject medical content absent from $\case$. Hand-tuning this prompt does not scale across the $3^7$ different configurations, so we instead optimize it on a small set of cases derived from a disjoint subset of VHA encounters~\citep{vha_2022} (Appendix~\ref{app:datasets}), held out from those used in the main experiments.

\paragraph{Dataset.}
Ten distinct patient profiles $\{\case^{(k)}\}_{k=1}^{10}$ are drawn from the $30$ held-out VHA encounters of Appendix~\ref{app:datasets}; the remaining $20$ held-out encounters are unused. For each profile, 30 patient configurations are generated with personality $\persona^{(k,j)} = (\persH,\persE,\persX,\persA,\persC,\persO,\persL)^{(k,j)} \in \{1,2,3\}$ sampled with uniform probability across all levels. The corpus of configurations is split \emph{by profile}: 7 profiles form the training pool and 3 form the held-out evaluation pool, so that no profile-specific phrasing leaks across the split. The optimizer itself operates on a fixed subsample of this split: $\mathcal{D}_{\mathrm{train}}$ (120 rows) drawn from the training pool and $\mathcal{D}_{\mathrm{eval}}$ (30 rows) drawn from the evaluation pool. Every row fixes the parameters $\convtype=\text{spoken}$ and $\answerformrate = 0.45$, so the tuned prompt is optimized under the same parametrization used for the main experiments.

\paragraph{Optimizer.}
We tune the prompt with CAPO~\citep{zehle_capo_2025}, an evolutionary prompt optimizer that alternates LLM-driven mutation and crossover steps, and eliminates candidates early via racing, as implemented in the \texttt{promptolution} framework~\citep{zehle_promptolution_2026}. We run CAPO for 30 steps with sequential-block evaluation, 20 subsamples per step, a length penalty of $0.05$, and no few-shot examples.

\paragraph{Reward.}
For each candidate prompt $\pi$ and each case-persona pair $(\case,\persona) \in \mathcal{D}$, we instantiate a patient whose latent role $\latentrole_0$ is generated by $\pi$ and run it against four questions designed to expose distinct trait combinations: (i)~\emph{symptom history and self-management} (eXtraversion, Emotionality); (ii)~\emph{lay causal explanation} (Openness, Emotionality); (iii)~\emph{worry elicitation} (Agreeableness, eXtraversion); (iv)~\emph{reassurance challenge} (Agreeableness, Openness). We then measure different penalties that determine the reward of $\pi$ on each sample: $P_{\mathrm{info}}(\pi,\case,\persona)$ is the fraction of hallucinated facts not in $\case$; $P_{\mathrm{pers}}(\pi,\case,\persona)$ is the mean absolute distance between the configured HEXACO levels and the levels reconstructed by the autorater from the responses, over the four axes the probes elicit (Emotionality, eXtraversion, Agreeableness, Openness), since Honesty-Humility and Conscientiousness act through the disclosure grid rather than through the latent role; $P_{\mathrm{pi}}(\pi,\case,\persona)$ is the share of personal-information fields missing from the generated role. Because $P_{\mathrm{pi}}$ is judged on the role \emph{text}, a longer role trivially covers more fields, so it is down-weighted relative to the behavioral penalties. Further, a penalty on the role's token length $|\latentrole_0|$ counteracts the resulting incentive to pad the role past what is needed:

\begin{align*}
R(\pi,\case,\persona) \;=\; -\Bigl[P_{\mathrm{info}} + P_{\mathrm{pers}} + 0.4\,P_{\mathrm{pi}} + 0.25\max\bigl(0,\, |\latentrole_0| - 1100\bigr) / 1100\Bigr]
\end{align*}   

As conversational LLM and autorater, we use \gemini{}, the same model used in the main experiments, so the optimization target is consistent with the main evaluation protocol.

\begin{figure}[h]
\centering
\includegraphics[width=0.75\textwidth]{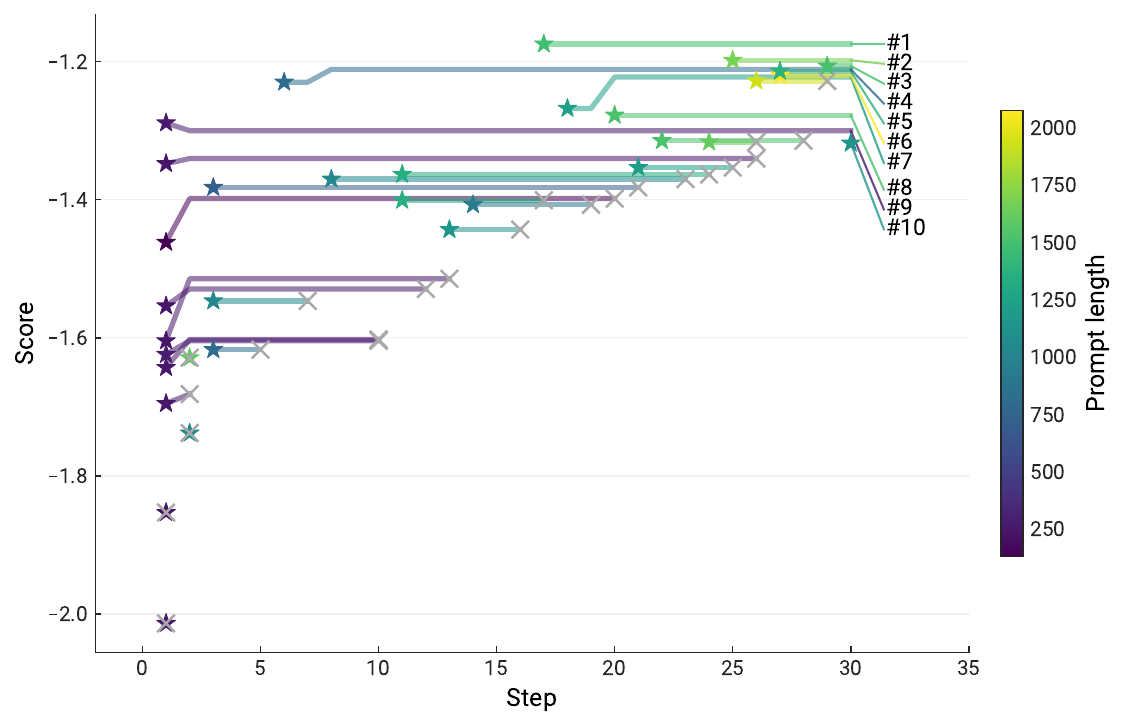}
\caption{Reward of every prompt in the population at the end of each step. The trajectory is colored by prompt length; stars mark the step at which a prompt entered the population and grey crosses mark elimination.}
\label{fig:metaprompt_step_scores}
\end{figure}

\paragraph{Results.}
Figure~\ref{fig:metaprompt_step_scores} shows the per-step score of every prompt surviving into the final population. Some unpromising initial candidates are pruned within the first few steps, and the surviving population's mean prompt length grows from 238 to 1{,}310 characters between step 1 and step 30 as its mean training-time score improves from $-1.61$ to $-1.23$, confirming that the population clearly converges over the 30 tuning steps. Evaluating the final population on $\mathcal{D}_{\mathrm{eval}}$ returns an unbiased estimate of the best prompt for our purpose. This prompt is used as $\platent$ for the main study. It organizes its instructions around five interactional dimensions (speech clarity, information sharing, affective expression, cognitive engagement, dialogue timing), each tied to concrete behavioral habits, under an explicit constraint to reproduce the provided HEXACO axes and personal details without modification.

\newpage
\section{Evaluation Study Details}\label{app:doctor_study_details}

The entire evaluation is based on the cases of the VHA dataset (cf. Appendix~\ref{app:datasets}). Throughout, the conversational model and the autorater are \gemini{} (temperature $1.0$, top-$p$ $0.95$, low thinking budget) and the meta-model is \mistral{} (temperature $0.05$). Nine resident clinicians annotated through a custom labeling interface in two independent blocks: a realism task and a personality task. Table~\ref{tab:default_configs} lists the simulator instances run on each case under the physician-aligned protocol.

\begin{table*}[h]
\centering
\footnotesize
\caption{Default simulator instances and parameters used in the realism comparison study. All instances generate their utterances with the same conversational model, \gemini{}.}
\label{tab:default_configs}
\begin{tabular}{lll}
\toprule
\textbf{Simulator} & \textbf{Parameters} & \textbf{Notes} \\
\midrule
Human Rephrase & — & Paraphrases ground-truth turns. \\
AgentClinic & no bias & Cognitive/social bias disabled. \\
\ours & $(2,1,1,2,2,1,\text{B})$ & Default persona. \\
PatientSim & B / plain / recall=high / dazed=normal & Default reference. \\
\bottomrule
\end{tabular}
\end{table*}

\subsection{Realism Study}\label{app:clinician_study_extended_results}

\paragraph{Annotation.}
Each clinician judged the same pool of 20 cases, every case shown as three side-by-side conversations rated independently. Exactly one of \emph{Human Reference} and \emph{Human Rephrase} appears per case, since the two render the same underlying turns; \ours appears on every case as the common reference point; the third column alternates between PatientSim and AgentClinic. Column order is shuffled per rater and per case, and columns carry no source labels: clinicians were told that any number of the three could be real and asked not to think of detecting the real one but rating the conversations independently. Marking a column not real required at least one reason, chosen from medically implausible, non-response, linguistically unnatural, inconsistent with prior context, too informative, uninformative, or free text. Because VHA encounters run between 43 and 98 turns, each column shows a short excerpt of the conversation. The excerpt is identical across the three columns of a case, fixed before any labels were collected, and chosen automatically as the window in which the real transcript discloses the most clinical detail, penalizing windows in which any simulator has entered a loop of saying goodbye multiple times. Since the penalty applies wherever any arm loops, it favors no simulator. All shown transcripts are ellipsis-normalized beforehand: the ellipses that the speech-to-text pipeline leaves in the recorded turns are replaced by commas or periods, uniformly across all patient types, so that a transcription artifact alone cannot mark a column as recorded. All other disfluencies are left untouched. This yields $540$ ratings: $180$ for \ours and $90$ for each of the four other patient types.

\begin{figure}[h]
\centering
\begin{minipage}[b]{0.49\linewidth}
\centering
\includegraphics[width=\linewidth]{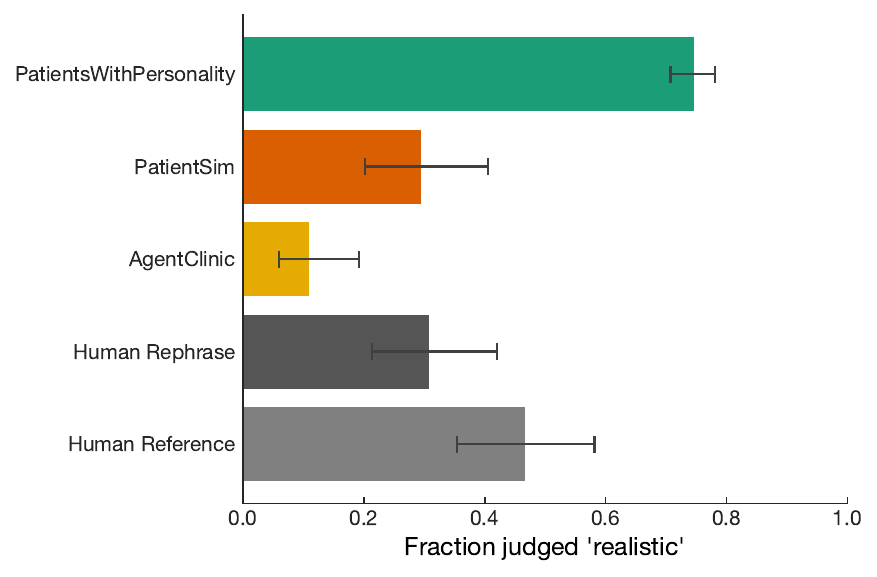}
\end{minipage}\hfill
\begin{minipage}[b]{0.49\linewidth}
\centering
\includegraphics[width=\linewidth]{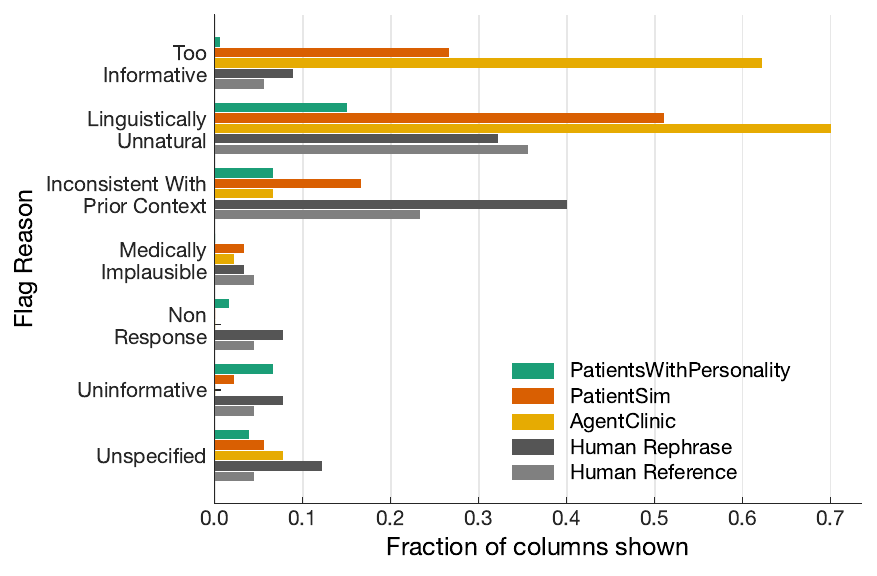}
\end{minipage}
\caption{Clinician ratings by patient type over the full nine-rater panel. \textbf{Left:} fraction of shown conversations rated realistic, as fitted by the mixed-effects logistic regression at the mean of both random effects, with $95\%$ intervals from the same fit. \textbf{Right:} reasons given when a conversation was marked not realistic, each as a fraction of all conversations shown for that patient type.}
\label{fig:realism_class_flags}
\end{figure}

\paragraph{Extended results.}
Figure~\ref{fig:realism_class_flags} reports the more detailed results Figure~\ref{fig:reason_scatter} is based on. The left panel reports the fraction judged realistic per patient type. Bars and intervals are read off the same Bayesian binomial mixed model with a fixed effect for patient type and crossed random intercepts for rater and case as used in Section~\ref{sec:results_default}. Each bar is the fitted rate at the mean of both random effects, and its interval covers $\pm 1.96$ posterior standard deviations of that fitted value on the logit scale. The fitted rates stay within $0.007$ of the observed fractions, and the estimated random-effect standard deviations, $0.42$ for rater and $0.25$ for case, are small next to the patient-type effects. The right panel gives every reason for classifying a conversation as `not real' as a fraction of all columns shown for that patient type.

\begin{figure}[h]
\centering
\includegraphics[width=0.5\linewidth]{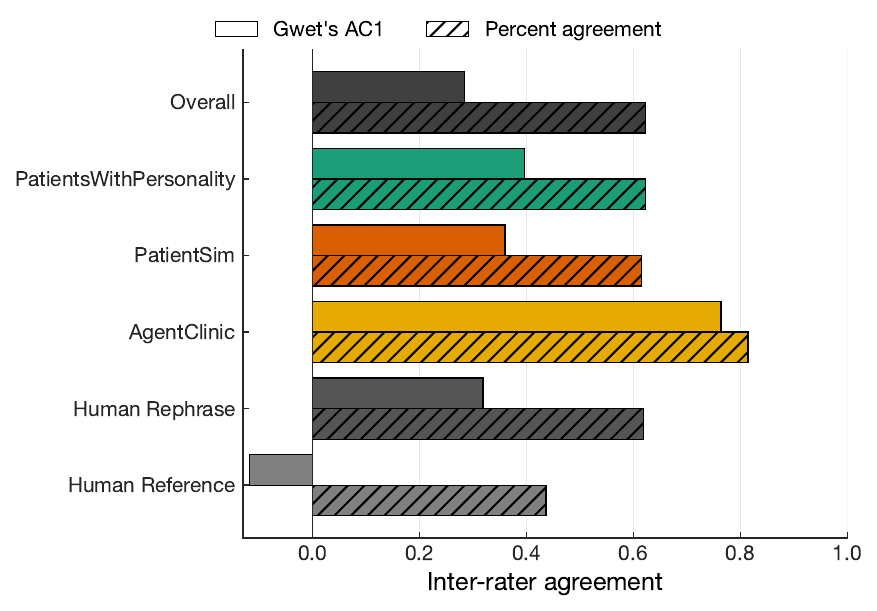}
\caption{Inter-rater agreement on the binary realistic/not-realistic rating over the nine-rater panel, overall and per patient type, as percent agreement and Gwet's $\mathrm{AC}_1$.}
\label{fig:interrater_agreement}
\end{figure}

\paragraph{Panel reliability.}
We compute agreement over the $98$ (case, patient type) items judged by at least two raters and report percent agreement together with Gwet's $\mathrm{AC}_1$~\citep{gwet_computing_2008}, which we prefer to Cohen's $\kappa$ because the marginals are strongly skewed (most conversations are rejected) and $\kappa$ collapses toward zero in that regime~\citep{feinstein_high_1990}. The panel reaches $62.3\%$ agreement with $\mathrm{AC}_1 = 0.28$ (cf. Figure~\ref{fig:interrater_agreement}).

While the agreement in general is moderate, there are some aspects where the raters disagree. The real recorded encounters split the panel completely (all $19$ are called realistic by some raters and not by others) yielding only $43.6\%$ agreement. This is most likely because the transcribed speech is full of disfluencies that some raters read as authentic and others as broken. Since $\mathrm{AC}_1$ depends on how often a patient type is called realistic at all, it is inflated for AgentClinic and deflated for the recorded encounters. We therefore never report the rating as an absolute detection rate, and every claim in Section~\ref{sec:results_default} compares patient types within a rater, which cancels individual thresholds.

\paragraph{Robustness of the realism comparison.}
Much of the disagreement stems from one rater who, contrary to the instructions, assumed the conversations were written chats and therefore penalized the fragmented flow of natural speech. Excluding this rater raises $\mathrm{AC}_1$ from $0.28$ to $0.38$, and the mixed-effects model still separates \ours from every other patient type. Among the remaining reasons, \emph{inconsistent with prior context} is the most frequent one given for the Human Reference (Figure~\ref{fig:realism_class_flags}). It arises in speech-to-text transcripts of in-person visits, where overlapping speech and cut-off turns are lost. Counting conversations rejected only for this reason as realistic raises the Human Reference by $19.4$ percentage points to $66.1\%$, the largest increase of all patient types, which narrows the gap to \ours.

\subsection{Extreme-Personality Study}\label{app:extreme_study}

\paragraph{Configuration.}
Each HEXACO axis is isolated by setting one component of $\persona$ to level $3$, its most challenging setting, while holding the other five at level $1$, their most cooperative setting, and fixing CEFR to B. Only PatientSim is included for comparison, as it is the only other tested simulator exposing explicit personality control. We pair each \ours configuration with the closest behavioral analogue in the PatientSim parameter space (cf. Table~\ref{tab:extreme_match}).

\begin{table*}[h]
\centering
\footnotesize
\caption{Pairing of extreme \ours configurations with the closest available PatientSim configuration.}
\label{tab:extreme_match}
\begin{tabular}{lll}
\toprule
\textbf{Persona} & \textbf{\ours $\persona$} & \textbf{PatientSim parametrization} \\
\midrule
Dishonest & $(3,1,1,1,1,1,\text{B})$ & B / distrust / recall=low / dazed=normal \\
Emotional & $(1,3,1,1,1,1,\text{B})$ & B / overanxious / recall=high / dazed=normal \\
Extraverted & $(1,1,3,1,1,1,\text{B})$ & B / verbose / recall=high / dazed=normal \\
Frustrated/Skeptical & $(1,1,1,3,1,1,\text{B})$ & B / impatient / recall=high / dazed=normal \\
Disorganized & $(1,1,1,1,3,1,\text{B})$ & B / plain / recall=low / dazed=high \\
Cautious & $(1,1,1,1,1,3,\text{B})$ & B / distrust / recall=high / dazed=normal \\
\bottomrule
\end{tabular}
\end{table*}

\paragraph{Diversity measurement.}
For the diversity analysis, the simulated patient turns of a conversation are concatenated and encoded into a single unit-norm vector with \qwen{}, and all reported distances are cosine distances between these vectors. Its $32$k-token context window holds the longest conversation of this study ($11.5$k tokens) in a single pass, so every conversation is encoded in full rather than truncated to its opening turns. Distances are always compared within a case.

\begin{figure}[h]
\centering
\includegraphics[width=0.45\linewidth]{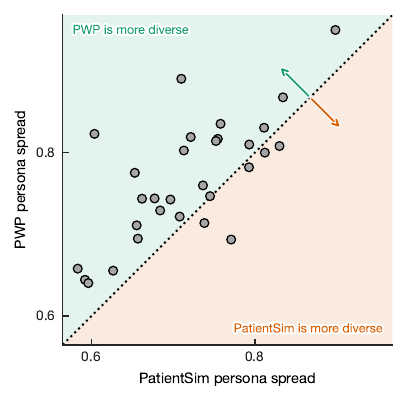}
\caption{Persona spread per clinical case for \ours and PatientSim, measured as the mean pairwise distance between the seven persona conversations generated for that case. Points above the diagonal, $25$ of the $30$ cases, are those on which \ours{} spans the wider behavioral range.}
\label{fig:case_spread}
\end{figure}

\paragraph{Annotation.}
Clinicians rated all six HEXACO axes from 1 to 3, or as \emph{unclear}, for each conversation shown. Each rater judged twelve conversations: per axis, one \ours configuration at that axis's extreme and its matched PatientSim configuration. The twelve were drawn from a rotating pool of seven candidate cases, six per rater without repetition and no case fixed across raters. Raters were not told the underlying parametrization, and in particular not that only a single axis was varied at a time. All nine completed the block, giving $108$ ratings over $62$ distinct conversation--simulator pairs.

\paragraph{Inter-rater agreement.}
The assignment was designed so that $30$ of the conversation--simulator pairs were rated independently by two or three clinicians. This gives $178$ rated traits and $352$ pairs of ratings. Krippendorff's $\alpha$~\citep{hayes_answering_2007} with an ordinal distance is $0.50$ (95\% bootstrap CI $[0.36, 0.61]$), a moderate agreement. $92\%$ of the rating pairs lie within one level on the 1--3 scale, and only $8\%$ place the same trait at opposite ends.

\paragraph{Independent autorater.}
\gemini{} generates the patient responses, serves as the autorater during prompt optimization (Appendix~\ref{app:metaprompt}), and scores the main personality study. To rule out that recovery reflects this shared model, we re-scored all conversations with \claude{}, which is involved neither in generation nor in prompt optimization, using the same autorater prompt. The two autoraters agree closely, with a Spearman correlation of $0.91$ on the mean recovered level per persona and axis.

\paragraph{Behavior at the extremes.}
The extreme configurations also cover patients who exaggerate or contradict themselves. With $\persE = \persO = 3$, simulated patients inflate symptoms beyond the case description (``worsening knee pain'' becomes ``cartilage is gone, can't even walk across the room without it locking up''), catastrophize, attribute unrelated problems to their self-diagnosis, and steer the visit back to their own complaints. With $\persC = 3$, medical facts start out omitted or fuzzy and only become precise under targeted questioning, so earlier answers can conflict with later ones. With $\persH = 3$, the patient is additionally instructed that it may give false information about its medical history (Table~\ref{tab:prompts-hexaco}).

\paragraph{Independent axes.}
The study varies one axis at a time, and the simulator permits any combination of levels, including ones that are rare in the population, so that robustness to such patients can be tested. The HEXACO domains are designed to be only weakly intercorrelated~\citep{ashton_empirical_2007}, so sampling them independently, as during prompt optimization, is a reasonable approximation; correlated sampling to match the population distribution is left to future work.

\section{Ablations of the Default Configuration}\label{app:ablations}

To test whether the realism results depend on the choice of the default configuration $(2,1,1,2,2,1,\text{B})$, we rerun three variants on the same $30$ cases under the physician-aligned protocol. \emph{Disclosure AUC} denotes the area under the disclosure-pacing curve (Figure~\ref{fig:disclosure_pacing}), where higher values indicate earlier disclosure; \emph{unprompted leakage} is the absolute rate underlying Table~\ref{tab:disclosure_metrics}.
\begin{itemize}
\item \textbf{All center} $(2,2,2,2,2,2,\text{B})$: The disclosure AUC barely changes ($0.302 \to 0.337$; Human Reference $0.333$), and unprompted leakage is even slightly lower with $\persX = 2$ ($0.116 \to 0.093$), so $\persX = 1$ does not favor the low-oversharing outcome. Only the style shifts, in the direction $\persX$ intends: readability moves closer to the Human Reference ($0.11 \to 0.61$; Human Reference $1.64$), but answers now exceed those of real patients by $8.0$ tokens, whereas the default falls short by only $2.3$. This supports $\persX = 1$ as the default for realistic, short patient turns.
\item \textbf{$\persA = 1$} $(2,1,1,1,2,1,\text{B})$: All metrics lie between the default and the all-center configuration; only the answer length differs from the default ($+1.4$ tokens).
\item \textbf{Disclosure grid off} (default persona, full case description in the prompt from the first turn): Unprompted leakage rises to the level of the other simulators ($0.116 \to 0.154$; PatientSim $0.151$), while the disclosure AUC ($0.331$) stays stable, since the latent role and the answer-form alignment still shape the pacing.
\end{itemize}
None of the claims in Section~\ref{sec:results_default} depends on the choice of the default configuration.

\section{Replication on ACI-Bench}\label{app:aci}

To test whether the results transfer beyond the VHA primary-care population, we compute the quantitative dialogue metrics and the disclosure AUC on $22$ cases of ACI-Bench~\citep{yim_aci-bench_2023}, sampled across specialties. ACI-Bench covers a broader population ($77\%$ female, aged 22--100) than the VHA veterans. Table~\ref{tab:aci} reports the difference to the dataset transcripts. \ours is closest on turn length, lexical diversity, and disclosure pacing, and its disclosure AUC matches the transcripts, while PatientSim and AgentClinic disclose faster on both datasets. \ours and PatientSim are comparable on domain terms, whereas AgentClinic uses about six times more clinical terms than \ours. PatientSim stays closer on readability, as on VHA (Figure~\ref{fig:sim_comp_metrics}).

\begin{table}[h]
\centering
\small
\caption{Difference to the ACI-Bench transcripts (mean $\pm$ SEM over conversations, $\text{sim}-\text{real}$). Bold marks the value closest to zero and any value within its SEM.}
\label{tab:aci}
\begin{tabular}{lccc}
\toprule
\textbf{Metric} & \textbf{\ours} & \textbf{PatientSim} & \textbf{AgentClinic} \\
\midrule
Token count $\Delta$ & $\mathbf{-0.74 \pm 1.45}$ & $+4.74 \pm 1.06$ & $+20.93 \pm 1.16$ \\
Lexical diversity $\Delta$ & $\mathbf{+0.00 \pm 0.02}$ & $-0.14 \pm 0.01$ & $-0.17 \pm 0.01$ \\
Domain term count $\Delta$ & $\mathbf{+12.3 \pm 7.4}$ & $\mathbf{+17.0 \pm 4.7}$ & $+77.5 \pm 7.8$ \\
Readability $\Delta$ & $+1.77 \pm 0.38$ & $\mathbf{+0.39 \pm 0.14}$ & $+5.75 \pm 0.19$ \\
Disclosure AUC $\Delta$ & $\mathbf{-0.01 \pm 0.02}$ & $+0.06 \pm 0.01$ & $+0.07 \pm 0.01$ \\
\bottomrule
\end{tabular}
\end{table}

\section{Contextual-Care Benchmark Details}\label{app:downstream}

\paragraph{Ground truth.}
The VHA transcripts come with Content Coding for Contextualization of Care (4C) annotations~\citep{vha_2022}: for each encounter, experts coded the problems that hinted at an underlying contextual factor (e.g., missed appointments, uncontrolled blood sugar), the factor itself (e.g., cost, transportation, health beliefs), and whether the physician's plan addressed it. For $92$ encounters with $206$ coded factors, each problem is added to the case description together with its factor, identically for all simulators.

\paragraph{Protocol.}
The doctor LLM receives only the clinician-observable cues of each problem, not its cause, and interviews the patient for at most $20$ turns. It is instructed to find out what in the patient's life is behind every flagged problem, asking about cost, access to care, transportation, understanding of the treatment, health beliefs, family or work demands, and emotional state, before deciding on a plan. Afterwards it states for each problem whether it identified a cause, the cause, and its plan. In the reference arm, the doctor LLM reads the real encounter instead of conducting an interview. We remove every turn that overlaps a coded care-plan span ($\approx 1.3\%$ of the text), so that the model cannot copy the physician's plan. All patients use \gemini{} as conversational model; \ours uses its default configuration, PatientSim the configuration of Table~\ref{tab:default_configs}, and AgentClinic no bias. \geminipro{} judges whether the stated cause matches the coded factor and whether the plan ignores it (contextual error). CIs are case-bootstrap 95\% intervals; differences between simulators use paired bootstraps over cases.

\paragraph{Results.}
Table~\ref{tab:downstream} lists the values shown in Figure~\ref{fig:downstream}.

\begin{table}[h]
\centering
\small
\caption{Contextual-care benchmark per doctor model. Values in parentheses are the difference to the real encounter; the last column is the mean absolute difference over the three models. Bold marks the patient type closest to the real encounter; $^{*}$ marks a significantly smaller distance than both PatientSim and AgentClinic (paired 95\% bootstrap CI over cases).}
\label{tab:downstream}
\begin{tabular}{lcccc}
\toprule
\textbf{Patient} & \gemini{} & \geminiflash{} & \claude{} & \textbf{Mean dist.} \\
\midrule
\multicolumn{5}{l}{\textit{Contextual error rate ($\downarrow$ better model or easier task)}} \\
\midrule
Real encounter & $0.45$ & $0.34$ & $0.15$ & -- \\
\ours & $\mathbf{0.33}\ (-0.13)^{*}$ & $\mathbf{0.23}\ (-0.11)^{*}$ & $\mathbf{0.15}\ (0.00)$ & $\mathbf{0.08}^{*}$ \\
PatientSim & $0.17\ (-0.28)$ & $0.12\ (-0.22)$ & $0.08\ (-0.07)$ & $0.19$ \\
AgentClinic & $0.20\ (-0.26)$ & $0.10\ (-0.24)$ & $0.07\ (-0.08)$ & $0.19$ \\
\midrule
\multicolumn{5}{l}{\textit{Cause correctly identified ($\uparrow$ better model or easier task)}} \\
\midrule
Real encounter & $0.63$ & $0.76$ & $0.86$ & -- \\
\ours & $\mathbf{0.58}\ (-0.05)$ & $\mathbf{0.69}\ (-0.07)$ & $0.80\ (-0.06)$ & $\mathbf{0.06}$ \\
PatientSim & $0.79\ (+0.16)$ & $0.86\ (+0.10)$ & $0.90\ (+0.04)$ & $0.10$ \\
AgentClinic & $0.77\ (+0.13)$ & $0.85\ (+0.10)$ & $\mathbf{0.89}\ (+0.03)$ & $0.09$ \\
\bottomrule
\end{tabular}
\end{table}

\paragraph{Example.}
In case \texttt{Z221}, the contextual factor is a health belief: the patient feels that medication should not ``go in'' the body. In the real encounter, the patient mentions this only in passing, while laughing, and all three models miss it. PatientSim states the belief plainly in its first answer, and every model addresses it (none of the nine runs commits a contextual error). \ours reproduces the real failure: the belief is buried among other reasons (``slips my mind\ldots{} pharmacy's a hassle\ldots{} don't like putting stuff in my system''). The model ends the interview after $5$ of $20$ turns, attributes the problem to forgetfulness, and plans an automatic refill and mail order, the same error it makes on the real encounter.

\paragraph{Personality configurations.}
In an exploratory run on $13$ coded factors with the three doctor models and three seeds, we replace the default persona with two extreme ones. A withholding patient ($\persH = \persC = 3$) reveals the factor in only $49\%$ of visits; correct cause identification drops by $0.32$ and the contextual error rises by $0.28$. An emotional patient with a strong prior belief ($\persE = \persO = 3$) still voices the factor in $84\%$ of visits, but misleads the model more often, e.g., with its self-diagnosis; correct cause identification drops by $0.15$ and the contextual error rises by $0.15$.

\section{Information Disclosure Analysis}\label{app:info_disclose_extended}

Table~\ref{tab:disclosure_metrics} reports disclosure as \emph{distance to the Human Reference}, since a low absolute disclosure level is not by itself realistic (a simulator could reach one by disclosing almost nothing). \emph{Pacing gap} is the mean absolute deviation between the disclosure-pacing curve (cf. Figure~\ref{fig:disclosure_pacing}) and the Human Reference over the first 30 turns. \emph{PDP F1 $\Delta$} is the gap in prompted-disclosure F1, which measures whether a field appears in the patient's answer when its content was explicitly requested; we reuse the relevant-field classifier $\Fmeta(\cdot;\pclassform)$ to identify the requested fields $\relevant_t$ at each turn. This is the same classifier that gates disclosure in \ours, so the metric is measured in the terms \ours optimizes and may favor it. The pacing gap, which does not use the classifier, does not share that dependence. \emph{Unprompted-leakage $\Delta$} is the gap in the rate at which fields surface in turns that did not request them. All three are differences of group means over the physician-aligned conversations, with bootstrap 95\% CIs. On that metric every simulator stays \emph{below} the Human Reference in absolute terms, so the larger distance of \ours reflects volunteering too little rather than oversharing. Because the rate normalizes by all disclosed facts, the smaller gap of the baselines is partly an artifact of their higher prompted disclosure:

\begin{table*}[h]
\centering
\small
\caption{Disclosure metrics by simulator, shown as distance to the Human Reference (with 95\% bootstrap CI). Smaller is closer to human disclosure behavior; the smallest value per column is marked bold.}
\label{tab:disclosure_metrics}
\footnotesize
\begin{tabular}{lccc}
\toprule
\textbf{Simulator} & \textbf{Pacing Gap $\downarrow$} & \textbf{PDP F1 $\Delta$ $\downarrow$} & \textbf{Unprompt.\ Leak.\ $\Delta$ $\downarrow$} \\
\midrule
PatientSim  & $0.195\ [0.162, 0.230]$ & $0.252\ [0.219, 0.285]$ & $\mathbf{0.006}\ [0.001, 0.052]$ \\
AgentClinic & $0.165\ [0.107, 0.229]$ & $0.246\ [0.212, 0.281]$ & $0.014\ [0.001, 0.069]$ \\
\midrule
\ours (\emph{Ours})     & $\mathbf{0.063}\ [0.031, 0.126]$ & $\mathbf{0.123}\ [0.092, 0.157]$ & $0.040\ [0.003, 0.092]$ \\
\bottomrule
\end{tabular}
\end{table*}

\end{document}